\documentclass[5p,times,lefttitle,sort&compress]{elsarticle}
\usepackage{bm}
\usepackage{xcolor}
\usepackage{lineno}
\modulolinenumbers[5]
\usepackage[colorlinks]{hyperref}
\usepackage{amsmath}
\usepackage[linesnumbered,boxed,vlined]{algorithm2e}
\newcommand{\rv}[1]{\textcolor{black}{#1}} 
\usepackage{float}
\usepackage{multirow}
\journal{Applied Energy (Accepted on April 19, 2021)}
\begin{document}

\begin{frontmatter}

\title{\rv{Blockchain-Based Decentralized Energy Management Platform for Residential Distributed Energy Resources in A Virtual Power Plant}}

\author[A]{Qing~Yang}
\ead{yang.qing@szu.edu.cn}
\author[B,C]{Hao~Wang\corref{mycorrespondingauthor}}

\cortext[mycorrespondingauthor]{Corresponding author.}
\ead{haowang6@stanford.edu}
\author[A]{Taotao~Wang}
\ead{ttwang@szu.edu.cn}
\author[A]{Shengli~Zhang}
\ead{zsl@szu.edu.cn}
\author[A]{Xiaoxiao~Wu}
\ead{xxwu.eesissi@gmail.com}
\author[A]{Hui~Wang}
\ead{whsz@szu.edu.cn}

\address[A]{College of Electronics and Information Engineering (CEI), Shenzhen University, Shenzhen, Guangdong Province, China}
\address[B]{Department of Data Science and AI, Faculty of Information Technology, Monash University, Melbourne, VIC 3800, Australia}
\address[C]{Stanford Sustainable Systems Lab, Stanford University, CA 94305, USA}

\begin{abstract}
The advent of distributed energy resources (DERs), such as distributed renewables, energy storage, electric vehicles, and controllable loads, \rv{brings} a significantly disruptive and transformational impact on the centralized power system. It is widely accepted that a paradigm shift to a decentralized power system with bidirectional power flow is necessary to the integration of DERs. The virtual power plant (VPP) emerges as a promising paradigm for managing DERs to participate in the power system. In this paper, we develop a blockchain-based VPP energy management platform to facilitate a rich set of transactive energy activities among residential users with renewables, energy storage, and flexible loads in a VPP. Specifically, users can interact with each other to trade energy for mutual benefits and provide network services, such as feed-in energy, reserve, and demand response, through the VPP. To respect the users' independence and preserve their privacy, we design a decentralized optimization algorithm to optimize the users' energy scheduling, energy trading, and network services. Then we develop a prototype blockchain network for VPP energy management and implement the proposed algorithm on the blockchain network. By experiments using real-world data trace, we validated the feasibility and effectiveness of our algorithm and the blockchain system. The simulation results demonstrate that our blockchain-based VPP energy management platform reduces the users' cost by up to 38.6\% and reduces the overall system cost by 11.2\%.
\end{abstract}


\end{frontmatter}


\section{Introduction}\label{sec:intro}

The fast-growing penetration of distributed energy resources (DERs), such as distributed renewables, energy storage, electric vehicles, and controllable loads, poses significant challenges to the centralized power systems with unidirectional power flow. Successful integration of heterogeneous DERs calls for a paradigm shift to a decentralized power system with bidirectional power flow. As such, virtual power plants (VPPs) attract considerable research attention as a promising paradigm to manage DERs. In the modern smart grid, the VPP aggregates the capacity of the heterogeneous DERs to form a cloud-based distributed power plant to provide grid services (e.g., feed-in energy, demand response, and ancillary service), as well as energy trading. Therefore, the VPP can replace the conventional power plant to achieve higher efficiency and better flexibility.

Currently, the operation of a VPP is often managed by a central coordinator who remotely controls all the VPP users. By this method, the coordinator collects a wide range of information from all the VPP users and processes them to generate the optimal energy schedule; then, the coordinator commands all the users to execute the optimal energy schedule in their VPP operation. However, as the number of VPP users increases, the conventional VPP energy management method faces three critical challenges. First, the operation of the VPP coordinator is a ``black box'' that cannot be verified and trusted by the VPP users. Second, the centralized VPP energy management collects all the users' private energy usage information, thus incurs privacy leakage concerns. Third, there is still a gap between the theory and the implementation of the VPP energy management method in practice. 

\subsection{Related works}
A large body of literature has focused on the energy management and scheduling of various DERs in VPPs. For example, a service-centric virtual power plant was studied in \cite{koraki2017wind} to integrate solar and wind energy generations into the electricity market by enabling the cooperation between the VPP and the distribution system operator. A similar work \cite{kasaei2017optimal} proposed to aggregate distributed generators, energy storage systems, and controllable loads in a VPP to mitigate the impact of the variable generations and uncertainties caused by wind and solar energy generations. The day-ahead scheduling of a renewable energy based virtual power plant was studied in \cite{zamani2016day} considering the uncertainties of market prices, electrical demand, and intermittent renewable power generation. A new VPP model was proposed in \cite{naval2020virtual} to integrate all available full-scale distributed renewable generations as a single plant and maximize its profit in the wholesale electricity market.
In addition to renewable energy resources in VPPs, many studies developed scheduling strategies for other energy resources, such as flexible demand and energy storage. For example, the authors in \cite{thavlov2014utilization} developed a method to use flexible demand in a building to defer power consumption from electric space heating controlled by an operational virtual power plant. The authors in \cite{royapoor2020building} demonstrated that deferrable loads, such as heat pumps, air handling units, lifts, lighting, circulating pumps, and dry air coolers, can be used to illustrate DR capability for a building VPP. In \cite{giuntoli2013optimized}, the authors proposed a new algorithm to optimize the day-ahead thermal and electrical scheduling of a VPP, including small-scale prosumers and energy storage.
A joint bidding strategy was proposed in \cite{tang2019optimal} to schedule energy storage systems, demand response, and renewable energy sources in VPPs for the maximum benefits in the energy market. A stochastic framework was developed in \cite{rashidizadeh124stochastic} for short-term scheduling of electric vehicle parking lots in a virtual power plant. Most of the above studies focused on energy management and scheduling in the energy market. 
\begin{figure*}[!ht]
    \centering
    \includegraphics[width=16cm]{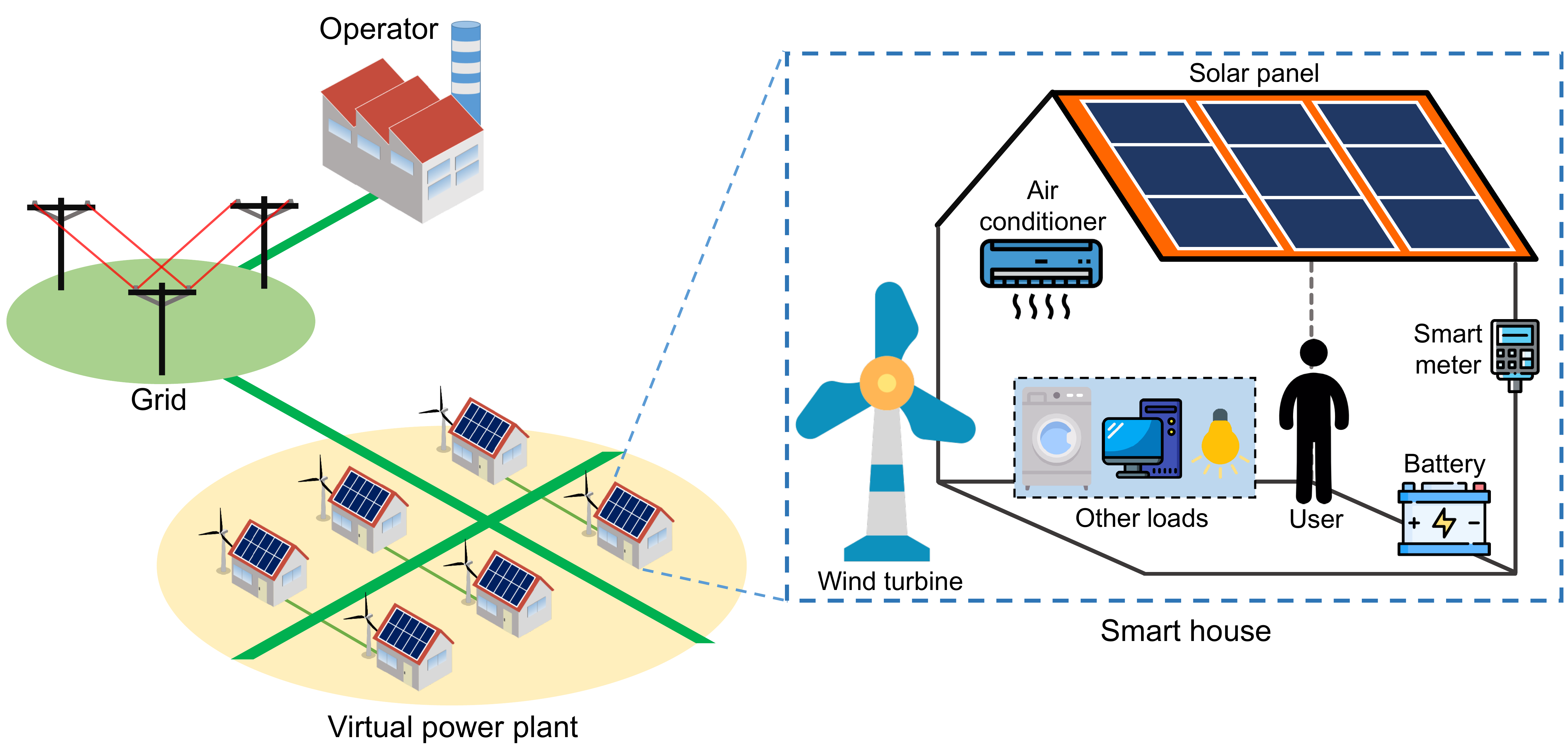}
    \caption{The system model of the virtual power plant that consists of multiple smart houses in the smart grid. The components of the smart house are elaborated in the right part of the figure.}
    \label{f1:sysmod}
\end{figure*}

More recent research explored various types of services that VPPs can provide and participate in the energy market. For example, in \cite{dabbagh2015risk, mashhour2010bidding, zhou2019four}, VPPs participated in a joint market of energy and reserve service, achieving higher profits. Energy trading as an emerging service was explored in \cite{nguyen2018optimizing, alam2019peer} 
In \cite{nguyen2018optimizing}, an optimization model was presented to maximize the economic benefits for  PV-battery distributed generations in a peer-to-peer (P2P) energy trading scenario. In \cite{alam2019peer}, the authors developed an algorithm to optimize the energy cost for smart homes under P2P energy trading and evaluated the impact of energy trading in a microgrid.
However, all the above studies adopted a centralized solution to the energy management and scheduling of VPPs, which is often not practical as DERs are often not owned by the VPP operator. 

Recent studies developed game-theoretic models \cite{wang2015interactive, hua2018stackelberg, yin2020energy} or decentralized solution methods to formulate more realistic scenarios for the operation of VPPs. For example, in \cite{wang2015interactive}, the market competition was modeled using game theory to determine the bidding strategy of VPPs, consisting of various distributed generations and battery storage devices. A Stackelberg game-theoretic model was developed in \cite{hua2018stackelberg} to capture the interaction between the operator and VPPs. In \cite{yin2020energy}, a robust Stackelberg game was proposed to manage aggregated prosumers in the form of VPP and participate in the day-ahead energy market. However, these game-theoretic models cannot be used to manage VPPs in real-world scenarios without an implementable algorithm. Therefore, implementable algorithms have been developed in a decentralized or distributed manner for energy management and scheduling of VPPs. Distributed algorithms based on the alternating direction method of multipliers (ADMM) were developed to facilitate energy trading among smart homes with heating, ventilation, and air conditioning (HVAC) \cite{yang2020cooperative} and interconnected microgrids \cite{wang2016incentivizing}, consisting of renewables, flexible loads, and energy storage. A decentralized optimization algorithm was proposed in \cite{li2016admm} to optimize the demand response of electric \rv{vehicles} in a VPP. A fully distributed algorithm was proposed in \cite{chen2018fully} using ADMM and consensus optimization for a VPP. Similarly, the study in \cite{dou124decentralized} proposed a decentralized aggregation strategy for multi-energy resources based on bi-level transactions of VPP. The above studies showed the great efforts made toward the distributed and decentralized energy management of distributed energy resources in VPPs. However, these method confronts two hurdles in practical implementation: first, the ADMM-based method still relies on a trusted central coordinator that incurs the risk of single-point failure; second, the process of the optimization method requires a trusted and verifiable computing environment that is hard to realize.

\rv{The word ``Blockchain'' first appears in the whitepaper of Bitcoin \cite{nakamoto2008bitcoin}, but its concept roots in the research area of distributed system \cite{natoli2019deconstructing}. The blockchain can be regarded as a distributed ledger that records the states of all the blockchain nodes \citep{dinh2018untangling}. For example, in Bitcoin, the state of a node is the bitcoin balance of the node, and the Bitcoin blockchain is a global ledger that stores the balances of all the nodes. The global state of \rv{the} decentralized ledger is maintained by and synchronized among all the nodes using the consensus protocol \cite{swan2015blockchain}. As a disruptive technology, blockchain prompts a paradigm shift in both academic and industrial areas of smart energy \rv{systems} recently \cite{hassan2019blockchain}.} LO3 Energy \cite{exergy} deployed a blockchain-based P2P energy trading platform named Exergy in the Brooklyn microgrid to facilitate online payments \cite{mengelkamp2018designing}. Exergy employed the blockchain technology only as a convenient payment tool for the users, but did not improve the efficiency of the trading system. In this work, we adopt the blockchain as a trustable computing machine to implement our energy management algorithm and also as a secure communication and payment tool.

Furthermore, the latest blockchain systems can execute smart contracts, e.g., Solidity on Ethereum \cite{eth}, allowing the users to implement generic programs on blockchain. \rv{Sabounchi \textit{et al.} \cite{sabounchi2017towards} used the Ethereum smart contract to implement a transactive energy trading algorithm based on auction theory. Wang \textit{et al.} proposed a P2P (peer-to-peer) energy crowdsourcing algorithm based on the Hyperledger smart contract in \cite{wang2019energy}. In both \cite{wang2019energy} and \cite{sabounchi2017towards}, the users' private information, including power consumption records and trading prices, are disclosed on the blockchain. By contrast, we design a decentralized energy management algorithm using the primal-dual method that preserves the users' privacy. To address the privacy issue, Li \textit{et al.} \cite{li2017consortium} designed a blockchain-based credit system to guarantee the privacy and security of the proposed transactive energy trading platform. The study in \cite{yang2021exploring} proposed the idea of using blockchain for the coordination of DERs, and \cite{yang2020blockchain} developed a blockchain-based transactive energy system to enable trustable energy trading among prosumers. However, [R6] relies on a centralized credit bank to manage the user's identity and credit information. Different from \cite{sabounchi2017towards, wang2019energy, li2017consortium}, our work considers a comprehensive decentralized VPP energy management platform without a central node, and we aim to implement the platform on practical smart meters. Different from \cite{yang2020blockchain}, our work considered both energy trading among DERs and grid services provided by DERs through the interaction between the DER aggregator and the grid.}

\subsection{Novelty and contribution}
In this work, we present a novel blockchain-based VPP energy management platform to address the above challenges. Blockchain is an open and verifiable distributed database that supports various cryptocurrency and decentralized applications (DApps). Furthermore, the blockchain is also a trustable computing machine that enables us to execute our energy management algorithm with the smart contract. By integrating blockchain technology, this work aims to develop an efficient, trustable, and privacy-preserving decentralized VPP energy management platform. The main contributions of this work are as follows.

\begin{enumerate}
  \item To the authors' best knowledge, this is the first work developing a trustable decentralized VPP energy management platform based on the blockchain technology that achieves correct and verifiable energy schedule.
  \item We consider a comprehensive set of transactive energy activities (such as energy trading and network services) for users and develop a distributed algorithm to optimize system efficiency and preserve the users' privacy.
  \item We elaborate on the design and implementation of the blockchain-based VPP energy management platform and build a prototype system to validate its effectiveness. 
\end{enumerate}

Outline of the manuscript: Section~\ref{sec:model} introduces the system model of the blockchain-based VPP energy management platform. Section~\ref{sec:problem} formulates the VPP energy management into mathematical problems. Section~\ref{sec:solution} elaborates on the design of the blockchian and the decentralized energy management algorithm for VPP. Section~\ref{sec:eval} implements a prototype system to evaluate the proposed platform with extensive experiments and Section~\ref{sec:conclusion} concludes our work.

\section{System Model}\label{sec:model}

This section describes the system model of the blockchain-based virtual power plant and the principle of the decentralized energy management. First, we present the model of the smart house with renewable energy generators, batteries, and different loads. Second, we introduce the virtual power plant that provides various services, including energy trading, demand response, feed-in tariff, and ancillary service. Third, we discuss the design and principle of the blockchain-based decentralized energy management system.

\subsection{The smart house}

With the growing penetration of renewable energy into the conventional power grid, the smart house becomes the prosumer that can not only consumes but also produces electricity. As shown in the right part of Fig.~\ref{f1:sysmod}, the smart house is equipped with renewable energy generators such as solar panels and wind turbines. It also supports various appliances (e.g., air conditioner, washer, lighting) to satisfy the user. Additionally, a battery energy storage system is installed in the smart house for storing extra electrical energy. Finally, these aforementioned components are managed and scheduled by the smart meter, which also connects the smart house to the local grid.

To aggregate the distributed energy resources (DER), a cluster of smart houses form a VPP via the existing power line as illustrated in Fig.~\ref{f1:sysmod}. We denote the smart house users that join the VPP by a set $u \in \mathcal{U} = \{1,\dots,U\}$, where $U$ is the total number of the VPP users. Without loss of generality, we assume the VPP schedules its service on the basis of a one-hour time slot. In this work, we adopt the day-ahead scheduling model as in \citep{nan2018optimal}, therefore the operational horizon is $t \in \mathcal{H} {=} \{1,\dots,24\}$ for the VPP system.

\begin{figure*}[!ht]
    \centering
    \includegraphics[width=16cm]{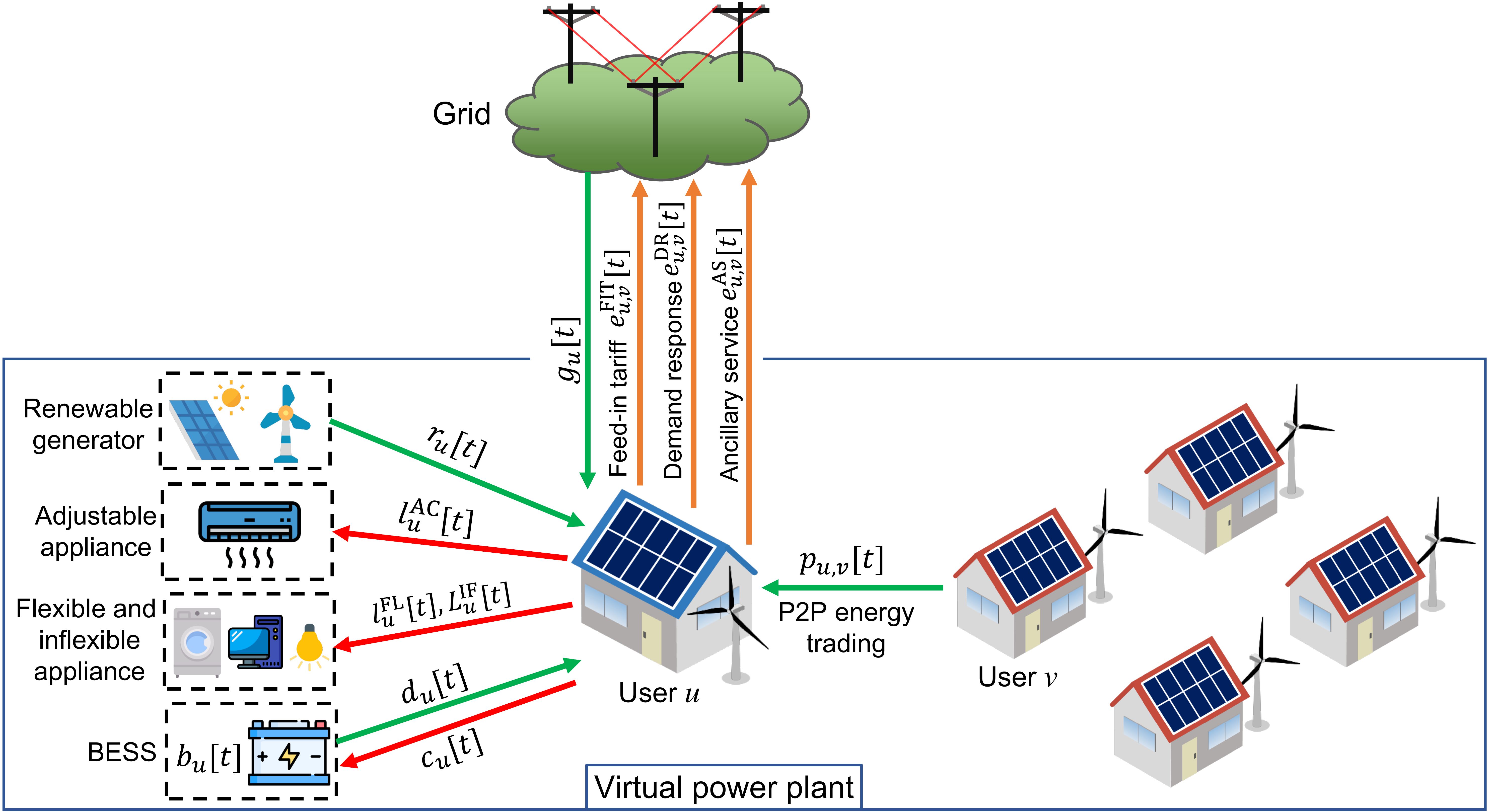}
    \caption{The principle and operational process of the virtual power plant.}
    \label{f2:nota}
\end{figure*}

\subsubsection{The power supply}\label{sec:supply}
The smart house user in the VPP can acquire electricity energy in several ways. First, the user can generate green energy using its renewable generators. Second, the user can buy electricity from the grid in the conventional way. Third, the user can also buy electricity from other VPP users via P2P energy trading. We next model the three sources of power supply, respectively.

Let $r_{t}^{u}$ denotes the amount of renewable energy generated by user $u$ at the  time slot $t$. The green energy is generated by the solar panel and wind turbine that varies with the natural environment such as the sunshine condition and wind speed; therefore, the renewable energy $r_{t}^{u}$ is upper-bounded by
    \begin{align}
        0 \leq r_{u}[t] \leq R_{u}[t], ~ \forall u \in \mathcal{U}, \forall t \in \mathcal{H},  \label{constraint-load5}
    \end{align}
where $R_{u}[t]$ is the amount of energy that user $u$'s renewable generator can produce during the time slot $t$. To emulate real-world green energy, we collect the power data from practical solar and wind generators installed at Hong Kong.

Since the users can always buy electricity from the grid, let $g_{u}[t]$ denotes the amount of electricity that user $u$ consumes from the grid in the $t$-th time slot. The grid power is limited by the electric fuse of the smart house, hence the following constraint applies
    \begin{equation}
        0 \leq  g_{u}[t] \leq G_{u}, ~ \forall u \in \mathcal{U}, \forall t \in \mathcal{H}, \label{constraint-load6}
    \end{equation}
where $G_{u}$ is the maximal input power allowed by the fuse system of user $u$. To improve peak shaving, the two-part tariff (TPT) pricing scheme is used by the grid operator to charge the grid users. Hence the user's payment to the grid operator is
    \begin{align}
            \mathcal{P}_u^{\mathrm{G}} = \alpha \sum_{t\in \mathcal{H}} g_{u}[t] + \beta \max_{t\in \mathcal{H}} g_{u}[t], \label{objective-supply} 
    \end{align}
where $\alpha$ is the normal electricity price and $\beta$ is the peak electricity price set by the grid operator. The peak price is higher than the normal price so that the users are encouraged to shave their peak loads.

To make full use of renewable energy, the user can trade electricity with other users of the VPP via P2P energy trading as proposed by \citep{alam2019peer, nguyen2018optimizing}. By allowing free energy trading, the users can effectively utilize the renewable energy and reduce their energy cost. Therefore, P2P energy trading is an important source of power supply for the VPP users. As shown in Fig.~\ref{f2:nota}, let $p_{u,v}[t]$ denotes the amount of energy that user $u$ purchased from user $v$ at the $t$-th time slot. Note here a positive $p_{u,v}[t]$ means user $u$ buys energy from user $v$, and a negative $p_{u,v}[t]$ means $u$ sells energy to user $v$. Therefore, the following constraints are established for energy trading
    \begin{align}
        p_{u,v}[t] & = -p_{v,u}[t], ~\forall t \in \mathcal{H},~\forall u,v \in \mathcal{U}, \label{constraint-trading1} \\
        p_{u,v}[t] & = 0~\textrm{if}~u = v, ~\forall t \in \mathcal{H}. \label{constraint-trading2}
    \end{align}
Note that Eq.~\eqref{constraint-trading1} guarantees that the energy trading market is balanced at each time slot, and Eq.~\eqref{constraint-trading2} avoids the invalid case in which the user trades with itself. The price of the P2P energy trading is set by the VPP operator to a fixed value  that is lower than the normal price of the grid. Let $\pi^{\mathrm{P2P}}$ denote the price of the P2P energy trading, then user $u$'s payment (or revenue) during the operational horizon is
\begin{equation}
    \mathcal{P}_u^{\mathrm{P2P}} = \sum_{t \in \mathcal{H}}  \sum_{v \in \mathcal{U}} \pi^{\mathrm{P2P}} p_{u,v}[t], \label{objective-tradingpayment}
\end{equation}
where a positive $\mathcal{P}_u^{\mathrm{P2P}}$ means user $u$'s overall payment to other users, and a negative $\mathcal{P}_u^{\mathrm{P2P}}$ means user $u$'s overall revenue from other users in the energy trading. \rv{The users can make payment conveniently using the token provided by the underlying blockchain system.} 

\subsubsection{The electric appliances}
As shown in the right part of Fig.~\ref{f1:sysmod}, various appliances are supported in the smart house. We categorize these appliances into three types: adjustable appliance, time-shiftable flexible appliance, and inflexible appliance. The adjustable appliance can be adjusted according to the smart house user's preference. For example, the air conditioner is a typical adjustable load because its power consumption depends on user's preferred temperature. The time-shiftable flexible appliance is the kind of loads that can be scheduled and shifted in any time slot of the operational horizon, such as the washer and dryer. Inflexible appliance (e.g., the house lighting and refrigerator) cannot be adjusted nor shifted in time.

In this work, we take the air conditioner as a typical example of the adjustable appliance. The function of the air conditioner is to keep the indoor temperature to the preferred value set by the user. Let $T_u[t]$ denote the indoor temperature of user $u$'s smart house at time $t$, and $\tau_u$ the preferred indoor temperature of user $u$. The indoor temperature depends on the power consumption of the air conditioner, denoted by $l^\mathrm{AC}_u[t]$, as well as the environmental temperature $T^{\mathrm{Out}}[t]$. Based on the dynamic model of air conditioner proposed in \citep{cui2019,lu2012evaluation}, the indoor temperature of user $u$ is
    \begin{align}
            T_{u}[t] = T^{\mathrm{Out}}[t] - \left( T^{\mathrm{Out}}[t] - T_{u}[t{-}1] \right) \mathrm{e}^{1/RC} + \gamma_u l^\mathrm{AC}_u[t{-}1], \nonumber\\
            \forall u \in \mathcal{U}, \forall t \in \mathcal{H}, \label{constraint-load1}
    \end{align}
where the coefficients $R$ and $C$ denotes respectively the equivalent heat capacity and thermal resistance of the air conditioner. The parameter $\gamma_u$ is the working mode and efficiency of user $u$'s air conditioner where a positive value indicates heating and a negative value indicates cooling.

We model the user $u$'s cost by the user's discomfort to the indoor temperature, which is defined as the deviation of the indoor temperature $T_u[t]$ to the user's preferred temperature $\tau_u$. Hence the user's discomfort cost is modeled by
    \begin{equation}
            \mathcal{P}_{u}^{\mathrm{AC}} = \omega_{\mathrm{AC}} \sum_{t \in \mathcal{H}} \left( T_{u}[t] - \tau_u \right)^{2}, ~ \forall u \in \mathcal{U}. \label{objective-load1}
    \end{equation}
In Eq.~\eqref{objective-load1}, we assign a weight coefficient $\omega_{\mathrm{AC}}$ to indicate the user's sensitivity to the discomfort term. In practice, the value of the indoor temperature should be in a reasonable range, therefore we have
\begin{equation}
    \underline{T}^{\mathrm{AC}} \le T_{u}[t] \le \overline{T}^{\mathrm{AC}}, \forall u \in \mathcal{U}, \forall t \in \mathcal{H}, \label{constraint-tmp}
\end{equation}
where $\underline{T}^{\mathrm{AC}}$ and $\overline{T}^{\mathrm{AC}}$ are the upper and lower bounds of the indoor temperature that can be achieved by the air conditioner.

We denote user $u$'s load of flexible appliance in time slot $t$ by $l_u^{\mathrm{FL}}[t]$. The flexible appliance can be scheduled to any time slot in the operational horizon $\mathcal{H}$ but should also follow two constraints. First, the gross sum of the flexible appliance should equate to the user's demand $L_u^{\mathrm{FL}}$. Second, the amount of flexible appliance per time slot should also be bounded in a reasonable range $[\underline{L}_u^{\mathrm{FL}}[t], \overline{L}_u^{\mathrm{FL}}[t]]$. Therefore, we obtain the following constraints for the flexible appliance
    \begin{align}
            \sum_{t \in \mathcal{H}} l_u^{\mathrm{FL}}[t] = L_u^{\mathrm{FL}}, &~ \forall u \in \mathcal{U}, \label{constraint-load3} \\
            \underline{L}_u^{\mathrm{FL}}[t] \leq  l_u^{\mathrm{FL}}[t] \leq  \overline{L}_u^{\mathrm{FL}}[t], &~ \forall u \in \mathcal{U}, \forall t \in \mathcal{H}. \label{constraint-load4}
    \end{align}
    
Usually the user $u$ has its preferred schedule of the flexible appliance, which is denoted by $L_u^{\mathrm{Ref}}[t], \forall t \in \mathcal{H}$. The preferred schedule reflects the user's most comfortable schedule of its flexible appliance, hence the deviation to the preferred schedule incurs a discomfort cost defined by
    \begin{align}
            \mathcal{P}_u^{\mathrm{FL}} = \omega_{\mathrm{FL}} \sum_{t \in \mathcal{H}} \left( l_u^{\mathrm{FL}}[t] - L_u^{\mathrm{Ref}}[t] \right)^{2}, \forall u \in \mathcal{U}, \label{objective-load2}
    \end{align}
where the coefficient $\omega_{\mathrm{FL}}$ indicates the user's sensitivity on the discomfort cost of the flexible appliance.

On the other hand, the inflexible appliances have regular (e.g., house lighting) or constant (e.g., refrigerators) power consumption that cannot be shifted over time. We denote the user $u$'s load of the inflexible appliances in the time slot $t$ by a sequence $L_u^{\mathrm{IF}}[t], \forall t \in \mathcal{H}$. Unlike the air conditioner and the flexible appliance, the inflexible appliance $L_u^{\mathrm{IF}}[t]$ must be supported but cannot be scheduled by the user.

\subsubsection{The battery energy storage system}
With the increasing penetration of renewable energy resources, the in-house battery energy storage system (BESS), e.g., the Tesla Powerwall, becomes popular in smart houses. The BESS can store the extra renewable energy in its high-capacity rechargeable battery and feed in energy to support appliances when needed. By aggregating the users' battery storage, the VPP can provide ancillary service to the grid.

We let $b_u[t]$ denote the energy charge level of user $u$'s battery in the $t$-th time slot. The amount of energy that is charged into and discharged from the battery at time slot $t$ are denoted by $c_u[t]$ and $d_u[t]$, respectively. Therefore, the charge level of the BESS is
    \begin{equation}
        b_u[t] = b_u[t-1] +  \eta c_u[t] - \frac{d_u[t]}{\eta}, t \in \mathcal{H}, \forall u \in \mathcal{U},  \label{constraint-battery1} \\
    \end{equation}
where the coefficient $\eta$ indicates the charging/discharging efficiency of the battery. The capacity and maximal power of the users' batteries are also bounded by their available resources and technical parameters, hence we obtain the following constraints
    \begin{align}
        0 \le b_u[t] \le \overline{B}_u, \forall t \in \mathcal{H}, \forall u \in \mathcal{U}, \label{constraint-battery2} \\
        0 \le c_u[t] \le \overline{C}_u, \forall t \in \mathcal{H}, \forall u \in \mathcal{U}, \label{constraint-battery3} \\
        0 \le d_u[t] \le \overline{D}_u, \forall t \in \mathcal{H}, \forall u \in \mathcal{U}, \label{constraint-battery4}
    \end{align}
where $\overline{B}_u$ is the capacity of the battery. The maximal charging and discharging power of user $u$'s battery are denoted by $\overline{C}_u$ and $\overline{D}_u$, respectively. The actual value of these parameters can be found in the specification of the BESS (e.g.,  $\overline{B}_u = 13.5$, $\overline{C}_u = 7$, and $\overline{D}_u = 7$ in kWh for the Tesla Powerwall~2 \citep{powerwall}).

Since charging and discharging degrade the lifespan of the battery, we model the user $u$'s cost on the BESS by
    \begin{align}
            \mathcal{P}_u^{\mathrm{BA}} = \omega_{\mathrm{BA}} \sum_{t\in \mathcal{H}} \left( c_u[t] + d_u[t] \right), \label{objective-battery}
    \end{align}
where $\omega_{\mathrm{BA}}$ is the coefficient that indicates the user's sensitivity to the battery cost.

\subsection{The virtual power plant}
In this work, we consider three grid services provided by the VPP: feed-in tariff (FIT), demand response (DR), and ancillary service (AS). The smart house users participate in the VPP by connecting their smart meters to the decentralized VPP management system, which is discussed in Section~\ref{sec:solution}. Unlike the conventional centralized VPP where the operator fully \rv{controls} the users' appliances, the proposed decentralized VPP only requires the users to share their trading decisions and lets the users schedule their appliances by themselves in a decentralized manner. 

\begin{figure*}[!ht]
    \centering
    \includegraphics[width=14cm]{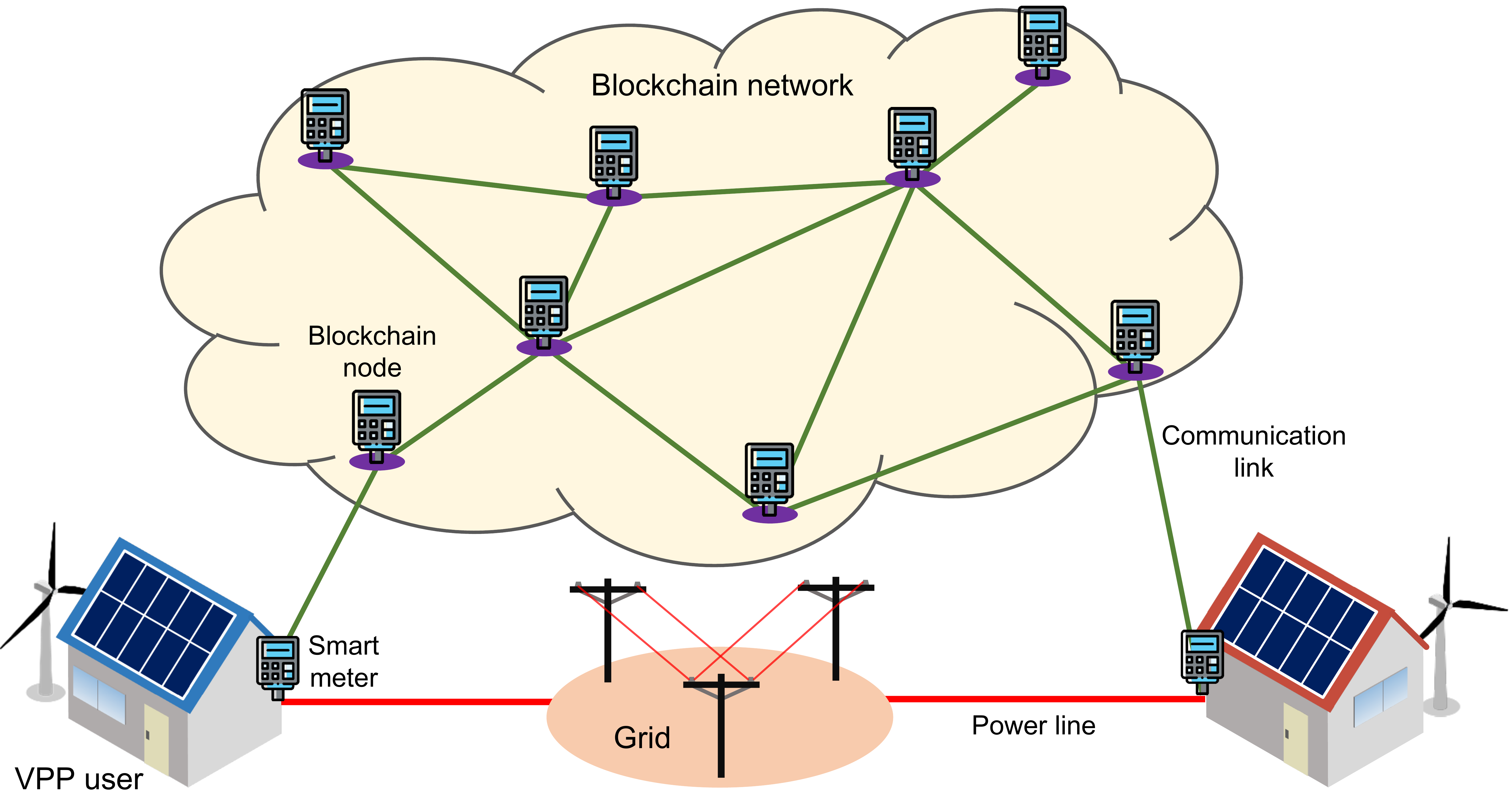}
    \caption{The blockchain system that supports the operation of the decentralized VPP energy management platform.}
    \label{f3:bc}
\end{figure*}

\subsubsection{The feed-in tariff}
The FIT service allows the VPP users to sell their renewable energy to the grid and receive the feed-in tariff as the monetary reward. The FIT mechanism accelerates the deployment of renewable generation by benefiting both the renewable energy producers and the grid. Without loss of generality, we assume the grid set a fixed FIT price denoted by $\pi^{\mathrm{FIT}}$ to the VPP users. Let $e_u^{\mathrm{FIT}}[t]$ denote the amount of user $u$'s FIT electricity at time slot $t$, then the user's reward is
\begin{align}
   \mathcal{R}_u^{\mathrm{FIT}} = \sum_{t\in \mathcal{H}} \pi^{\mathrm{FIT}} e_u^{\mathrm{FIT}}[t]. \label{objective-fit} 
\end{align}
The FIT is also bounded by the available renewable energy generation of the user, therefore we have the following constraints
\begin{align}
    e_u^{\mathrm{FIT}}[t] \leq R_{u}[t] - r_{u}[t], ~\forall u \in \mathcal{U}, t \in \mathcal{H}, \label{constraint-load9}
\end{align}
where the right hand side is the user's remaining energy that can be used as FIT.

\subsubsection{The demand response}

The demand response mechanism motivates the VPP users to adjust their power consumption demand to match the grid supply with certain rewards. The grid operator can send the DR request to signal the users to reduce their power consumption during the peak hours (usually in late afternoon and evening). To respond the DR signal, the user can reduce their electricity supply from the grid by adjusting its schedule of the adjustable appliances and flexible appliances. Let $e_u^{\mathrm{DR}}[t]$ \rv{denote} the user $u$'s reduction of the grid electricity usage in the $t$th time slot. Then the user can receive the DR reward $\mathcal{R}_u^{\mathrm{DR}}$ from the grid operator such that
\begin{align}
   \mathcal{R}_u^{\mathrm{DR}} = \sum_{t \in \mathcal{H}} \pi^{\mathrm{DR}}[t] e_u^{\mathrm{DR}}[t], \label{objective-dr} 
\end{align}
where $\pi^{\mathrm{DR}}[t]$ is the DR price set by the grid operator. Note here the operator can set different DR prices for different operational time slots, e.g., a higher DR price for peak hours to achieve peak shaving. Also, the user's amount of DR response is bounded by its scheduled grid electricity usage $g_u[t]$; hence we have the following constraint
\begin{align}
    0 \leq e_u^{\mathrm{DR}}[t] \leq g_u[t], &~ \forall u \in \mathcal{U}, t \in \mathcal{H}. \label{constraint-load10} 
\end{align}

\subsubsection{The ancillary service}
The VPP provides ancillary service to the grid by utilizing the users' battery storage capability. The VPP pay the users with the AS rewards to incentivize them to reserve some energy in their BESS for load regulation or spinning reserves. The reserved energy can be immediately dispatched from the VPP to the grid to maintain the grid stability or help the grid recover  \citep{melo2016robust}. Let $e_u^{\mathrm{AS}}[t]$ denotes the amount of energy reserved by user $u$ in its BESS at the $t$-th time slot for the AS. Then the total reward this user can earn is
\begin{align}
   \mathcal{R}_u^{\mathrm{AS}} = \sum_{t \in \mathcal{H}} \pi^{\mathrm{AS}}[t] e_u^{\mathrm{AS}}[t], \label{objective-as} 
\end{align}
where $\pi^{\mathrm{AS}}[t]$ is the unit price of AS at the $t$-th time slot. The grid operator can set this price according to the forecasted grid demand to ensure a stable operation. On the other hand, the amount of user's AS is bounded by its battery energy level $b_u[t]$. Therefore, we have the following constraint
\begin{align}
    0 \leq \rv{e_u^{\mathrm{AS}}[t]} \leq b_u[t], &~ \forall u \in \mathcal{U}, t \in \mathcal{H}. \label{constraint-as} 
\end{align}

\subsection{The blockchain system for the VPP energy management}

In this work, we employ the blockchain technology to support the implementation of the VPP energy management platform to achieve three goals. First, we build an open and verifiable energy management platform for the VPP. Unlike the conventional centralized VPP management method, the blockchain is a trusted computing machine that can run the energy management algorithm, thus \rv{removes} the need for a central coordinator. Second, the blockchain provides a secure and robust communication network. Third, the blockchain's digital currency provides a useful payment tool for the energy trading and rewards of network services.

We let the users' smart meters \rv{join} the blockchain network as the blockchain nodes, as shown in Fig.~\ref{f3:bc}. Current smart meters are embedded smart devices that can deal with complicated computing tasks \citep{weranga2014smart}. Running the blockchain node on the smart meter makes use of the existing smart meters. In Fig.~\ref{f3:bc}, the smart meters connect to the grid by the powerline and connect to the blockchain network by the communication link such as LoRa and 5G Narrowband IoT. These connected smart meters form a peer-to-peer network to transmit blockchain messages via the gossip protocol.

The consensus protocol is used by the blockchain nodes to synchronize their local state with other nodes in the network. The consensus protocol is a crucial component that affects the overall performance of the blockchain system. In this work, we adopt the proof-of-authority (PoA) consensus protocol rather than the popular proof-of-work (PoW) because PoW consumes too much computational resource that a smart meter cannot afford. For example, running a Bitcoin full-nodes requires at least 200GB disk space, 2GB memory, 200Kbps network bandwidth, and a CPU that can support a recent version of the operating system \cite{bitcoinfull}.

By contrast, the computational complexity of PoA consensus protocol is low, so that it is feasible on smart meters. Furthermore, PoA has a much shorter transaction confirmation time, which is important for network services through the VPP. In the PoA consensus protocol, a group of \emph{PoA nodes} \rv{is} selected as \rv{the PoA committee} to participate \rv{in} the consensus protocol and generate blocks. Other smart meters are \emph{normal nodes} that can send transactions but do not generate the block. The PoA nodes are responsible for receiving the transactions, executing them, and packaging them into a new block. \rv{The PoA node can also propose to add a new node into or remove an exiting PoA node from the committee, and let other PoA nodes vote on its proposal. If more than half of the PoA nodes agree on the proposal, then the proposal get executed and the members of the PoA committee change accordingly. This democratic voting mechanism guarantees the decentralization of the PoA-based blockchain.}

To support VPP energy management, the blockchain offers three types of transactions. The first type is the network service transaction that carries the FIT information $\bm{e}_u^{\mathrm{FIT}}$, DR information  $\bm{e}_u^{\mathrm{DR}}$, and AS information $\bm{e}_u^{\mathrm{AS}}$. This type of transaction is made between the VPP users and the grid. The second type is the P2P energy trading transaction that carries the trading information ${p}_{u,v}[t]$. This type of transaction is made by VPP users to interact with the VPP energy management platform. The third type is the token transfer transaction, \rv{since the blockchain provides the token as a digital currency (like Bitcoin) to ease the online payment. The token transfer transaction allows the users to pay for P2P energy trading and the VPP operator to pay the rewards.} The detailed implementation of the blockchain system will be elaborated later in Section~\ref{s:alg}.

In this section, we described the system model of the blockchain-based VPP system shown in Fig.~\ref{f1:sysmod}. A brief summary of the VPP's operational process and mathematical notations is illustrated in Fig.~\ref{f2:nota}. The problem of decentralized energy management for the VPP will be discussed in the next section.

\section{Problem Formulation of the Virtual Power Plant}\label{sec:problem}

This section formulates the working principle of VPP into mathematical problems. We first consider a benchmark VPP platform where the users work in a \emph{standalone mode} that each user individually schedules its usage of appliances to provide network services. We then consider the proposed VPP system where the users work in a \emph{cooperative mode} that the users can trade electricity and exchange their trading information to increase the profit of the VPP.

\subsection{The standalone (SA) mode VPP}

In the conventional centralized VPP, each user individually schedules their energy usage and interacts independently with the grid to provide the network services. Therefore, in this mode each user tries to maximize its local reward by managing its power supplies (from both grid and renewable generators), electric appliances (adjustable, flexible, and inflexible appliances), battery, and network services (FIT, DR, and AS).

During the VPP management, the power supply and consumption shall always be balanced in each user's smart house; therefore, the following constraint must be satisfied during the operational horizon
\begin{equation}
        \begin{aligned}
            l_u^{\mathrm{AC}}[t] + l_u^{\mathrm{FL}}[t] + L_u^{\mathrm{IF}}[t] + c_u[t] +  e_u^{\mathrm{DR}}[t]  = r_u[t] + g_u[t] &\\
            + d_u[t], \forall t \in \mathcal{H}, \forall u \in \mathcal{U}, & \label{constraint-SA}
        \end{aligned}
\end{equation}
where the L.H.S. sums up all the power consumed by the adjustable appliance $l_u^{\mathrm{AC}}[t]$, the flexible appliance $l_u^{\mathrm{FL}}[t]$, inflexible appliances $L_u^{\mathrm{IF}}[t]$, battery charging $c_u[t]$, and demand response $e_u^{\mathrm{DR}}[t]$. The R.H.S. of \eqref{constraint-SA} sums up all the power supplies including the renewable energy $r_u[t]$, energy from the grid $g_u[t]$, and battery discharging $d_u[t]$.

\begin{table}[!t]
    \centering
    \caption{The definitions of the vector variables.}\label{tab:short}
    \footnotesize
    \renewcommand{\arraystretch}{1.3}
    \label{t1:cost}
    \begin{tabular}{c c l}
        \hline
        \textbf{Variable} & \textbf{Definition} & \textbf{Meaning} \\
        \hline
        $\bm{l}_u^{\mathrm{AC}}$ & $\langle l_u^{\mathrm{AC}}[1], \dots, l_u^{\mathrm{AC}}[24] \rangle^{T}$ & User $u$'s adjustable appliance \\
        \hline
        $\bm{l}_u^{\mathrm{FL}}$ & $\langle l_u^{\mathrm{FL}}[1], \dots, l_u^{\mathrm{FL}}[24] \rangle^{T}$ & User $u$'s flexible appliance \\
        \hline
        $\bm{L}_u^{\mathrm{IF}}$ & $\langle L_u^{\mathrm{IF}}[1], \dots, L_u^{\mathrm{IF}}[24] \rangle^{T}$ & User $u$'s inflexible appliance \\
        \hline
        
        $\bm{r}_u$ & $\langle r_u[1], \dots, r_u[24] \rangle^{T}$ & User $u$'s renewable generation \\
        \hline
        $\bm{p}_{u,v}$ & $\langle p_{u,v}[1], \dots, p_{u,v}[24] \rangle^{T}$ & Energy trading between user $u$ and $v$ \\
        \hline
        $\bm{g}_u$ & $\langle g_u[1], \dots, g_u[24] \rangle^{T}$ & User $u$'s grid usage \\
        \hline
        $\bm{c}_u$ & $\langle c_u[1], \dots, c_u[24] \rangle^{T}$ & User $u$'s battery charging\\
        \hline
        $\bm{d}_u$ & $\langle d_u[1], \dots, d_u[24] \rangle^{T}$ & User $u$'s battery discharging \\
        \hline
        $\bm{e}_u^{\mathrm{DR}}$ & $\langle e_u^{\mathrm{DR}}[1], \dots, e_u^{\mathrm{DR}}[24] \rangle^{T}$ & User $u$'s demand response  \\
        \hline
        $\bm{e}_u^{\mathrm{FIT}}$ & $\langle e_u^{\mathrm{FIT}}[1], \dots, e_u^{\mathrm{FIT}}[24] \rangle^{T}$ & User $u$'s \rv{feed-in energy}\\
        \hline
        $\bm{e}_u^{\mathrm{AS}}$ & $\langle e_u^{\mathrm{AS}}[1], \dots, e_u^{\mathrm{AS}}[24] \rangle^{T}$ & User $u$'s \rv{auxiliary service} \\
        \hline
    \end{tabular}
\end{table}

To make the notation in \eqref{constraint-SA} more concise, we rewrite all the variables in vector form as defined in Table~\ref{tab:short}. Then, the constraint of \eqref{constraint-SA} can be rewrite as
\begin{equation}
    \bm{l}_u^{\mathrm{AC}} + \bm{l}_u^{\mathrm{FL}} + \bm{L}_u^{\mathrm{IF}} + \bm{c}_u + \bm{e}_u^{\mathrm{DR}} = \bm{r}_u + \bm{g}_u + \bm{d}_u, ~ \forall u \in \mathcal{U}. \label{constraint-load11}
\end{equation}

User $u$'s operational cost in the standalone mode VPP is its overall payment minus its overall rewards, then 
\begin{align}
             &\mathcal{P}^{\mathrm{SA}}_u =   \mathcal{P}_u^{\mathrm{G}} + \mathcal{P}_u^{\mathrm{AC}} + \mathcal{P}_u^{\mathrm{FL}} + \mathcal{P}_u^{\mathrm{BA}} - \mathcal{R}_u^{\mathrm{FIT}} - \mathcal{R}_u^{\mathrm{DR}} - \mathcal{R}_u^{\mathrm{AS}} \nonumber\\
             &= \alpha \sum_{t\in \mathcal{H}} g_{u}[t] + \beta \max_{t\in \mathcal{H}} g_{u}[t] + \gamma_{\mathrm{b}} \omega_{\mathrm{AC}} \sum_{t \in \mathcal{H}} \left( T_{u}[t] {-} \tau_u \right)^{2} \nonumber\\
             &{+} \omega_{\mathrm{FL}} \sum_{t \in \mathcal{H}} \left( l_u^{\mathrm{FL}}[t] {-} L_u^{\mathrm{Ref}}[t] \right)^{2} {+} \omega_{\mathrm{BA}} \sum_{t\in \mathcal{H}} \left( c_u[t] {+} d_u[t] \right) \nonumber\\
             &{-} \sum_{t\in \mathcal{H}} \pi^{\mathrm{FIT}} e_u^{\mathrm{FIT}}[t] {-} \sum_{t \in \mathcal{H}} \pi^{\mathrm{DR}}[t] e_u^{\mathrm{DR}}[t] {-} \sum_{t \in \mathcal{H}} \pi^{\mathrm{AS}}[t] e_u^{\mathrm{AS}}[t], \label{objective-operatingcost} 
\end{align}
where $\mathcal{P}^{\mathrm{SA}}_u$ denotes user $u$'s operational cost in the standalone mode. Each VPP user $u$ tries to minimize $\mathcal{P}^{\mathrm{SA}}_u$ by scheduling its power supplies (grid supply $\bm{g}_u$, renewable generation $\bm{r}_u$), electric appliances (adjustable appliance $\bm{l}_u^{\mathrm{AC}}$, flexible appliance $\bm{l}_u^{\mathrm{FL}}$), battery system ( charging $\bm{c}_u$ and discharging $\bm{d}_u$), and network services (feed-in tariff $\bm{e}_u^{\mathrm{FIT}}$, demand response $\bm{e}_u^{\mathrm{DR}}$, and ancillary service  $\bm{e}_u^{\mathrm{AS}}$). Therefore, the energy management problem of the VPP can be formulated into the following optimization problem
    \begin{equation}
        \begin{aligned}
            \bm{s}_u^{\mathrm{SA}} = 
            & \arg \min_{\bm{g}_u, \bm{r}_u, \bm{l}_u^{\mathrm{AC}}, \bm{l}_u^{\mathrm{FL}}, \bm{c}_u, \bm{d}_u, \bm{e}_u^{\mathrm{FIT}}, \bm{e}_u^{\mathrm{DR}}, \bm{e}_u^{\mathrm{AS}}} \mathcal{P}^{\mathrm{SA}}_u,  \label{opt1}\\
            &\mathrm{subject} \: \mathrm{to} \: \mathrm{constraints} \: 
            \mathrm{\eqref{constraint-load5}},
            \mathrm{\eqref{constraint-load6}},
            \mathrm{\eqref{constraint-load1}},
            \mathrm{\eqref{constraint-tmp}},
            \mathrm{\eqref{constraint-load3}},
            \mathrm{\eqref{constraint-load4}},\\
            & \mathrm{\eqref{constraint-battery1}}-\mathrm{\eqref{constraint-battery4}},
            \mathrm{\eqref{constraint-load9}},
            \mathrm{\eqref{constraint-load10}},
            \mathrm{\eqref{constraint-as}},
            \mathrm{\eqref{constraint-load11}},
        \end{aligned}
    \end{equation}
where the solution $\bm{s}_u^{\mathrm{SA}} {\triangleq} (\bm{g}^{\star}_u, \bm{r}^{\star}_u, \bm{l}^{\mathrm{AC} \star}_u, \bm{l}^{\mathrm{FL} \star}_u, \bm{c}^{\star}_u, \bm{d}^{\star}_u, \bm{e}^{\mathrm{FIT} \star}_u, \bm{e}^{\mathrm{DR} \star}_u, \bm{e}^{\mathrm{AS} \star}_u)$ represents the user $u$'s optimal energy schedule in the SA mode. Because the optimization objective function is convex as shown in \eqref{objective-operatingcost}, the user can locally solve the optimization problem of \eqref{opt1} with standard convex optimization tools. Therefore, we omit the solution of the above optimization problem, and let its solution $\bm{s}_u^{\mathrm{SA}}$ be the benchmark result to compare with the cooperative mode VPP discussed in the next section.

\subsection{The cooperative (CO) mode VPP}

In the this mode, the users cooperatively schedule their in-house appliances and network services by exchanging surplus energy with each other. In this mode, the energy trading must be considered in the constraint, so the original constraint in Eq.~\eqref{constraint-load11} is updated to
\begin{equation}
    \bm{l}_u^{\mathrm{AC}} + \bm{l}_u^{\mathrm{FL}} + \bm{L}_u^{\mathrm{IF}} + \bm{c}_u + \bm{e}_u^{\mathrm{DR}} = \bm{r}_u + \bm{g}_u + \bm{d}_u + \sum_{v \in \mathcal{U} \backslash u} \bm{p}_{u,v}, ~ \forall u \in \mathcal{U}, \label{constraint-load12}
\end{equation}
where the new term $\sum\nolimits_{v \in \mathcal{U} \backslash u} \bm{p}_{u,v}$ represents the total electricity that user $u$ purchases from other VPP users via the P2P energy trading as described in Section~\ref{sec:supply}. In the cooperative mode, the users' payments of the energy trading are also included in their operational costs, hence user $u$'s operational cost is
\begin{align}
             \mathcal{P}^{\mathrm{CO}}_u = & \mathcal{P}_u^{\mathrm{G}} + \mathcal{P}_u^{\mathrm{AC}} + \mathcal{P}_u^{\mathrm{FL}} + \mathcal{P}_u^{\mathrm{BA}} + \mathcal{P}_u^{\mathrm{P2P}} - \mathcal{R}_u^{\mathrm{FIT}} - \mathcal{R}_u^{\mathrm{DR}} - \mathcal{R}_u^{\mathrm{AS}} \nonumber\\
             = & \alpha \sum_{t\in \mathcal{H}} g_{u}[t] + \beta \max_{t\in \mathcal{H}} g_{u}[t] + \gamma_{\mathrm{b}} \omega_{\mathrm{AC}} \sum_{t \in \mathcal{H}} \left( T_{u}[t] {-} \tau_u \right)^{2} \nonumber\\
             &{+} \omega_{\mathrm{FL}} \sum_{t \in \mathcal{H}} \left( l_u^{\mathrm{FL}}[t] {-} L_u^{\mathrm{Ref}}[t] \right)^{2} {+} \omega_{\mathrm{BA}} \sum_{t\in \mathcal{H}} \left( c_u[t] {+} d_u[t] \right) \nonumber\\ 
             &+ \sum_{t \in \mathcal{H}}  \sum_{v \in \mathcal{U}} \pi^{\mathrm{P2P}} p_{u,v}[t] 
             {-} \sum_{t\in \mathcal{H}} \pi^{\mathrm{FIT}} e_u^{\mathrm{FIT}}[t] {-} \sum_{t \in \mathcal{H}} \pi^{\mathrm{DR}}[t] e_u^{\mathrm{DR}}[t] \nonumber\\  
             &{-} \sum_{t \in \mathcal{H}} \pi^{\mathrm{AS}}[t] e_u^{\mathrm{AS}}[t]. \label{objective-operatingcost2} 
\end{align}

Accordingly, the user $u$'s energy management is to obtain the optimal energy schedule $\bm{s}_u^{\mathrm{CO}}$ that solves the following optimization problem:
    \begin{equation}
        \begin{aligned}
            \bm{s}_u^{\mathrm{CO}} = & \arg \min_{\bm{g}_u, \bm{r}_u, \bm{l}_u^{\mathrm{AC}}, \bm{l}_u^{\mathrm{FL}}, \bm{c}_u, \bm{d}_u, \bm{p}_{u,v \in \mathcal{U}}, \bm{e}_u^{\mathrm{FIT}}, \bm{e}_u^{\mathrm{DR}}, \bm{e}_u^{\mathrm{AS}}} \mathcal{P}^{\mathrm{CO}}_u, \label{opt2}\\
            & \mathrm{subject} \: \mathrm{to} \:  \mathrm{constraints} \:  \mathrm{\eqref{constraint-load5}},
            \mathrm{\eqref{constraint-load6}},
            \eqref{constraint-trading1}, \eqref{constraint-trading2},
            \mathrm{\eqref{constraint-load1}},
            \mathrm{\eqref{constraint-tmp}},
            \mathrm{\eqref{constraint-load3}},
            \mathrm{\eqref{constraint-load4}},\\
            & \mathrm{\eqref{constraint-battery1}}-\mathrm{\eqref{constraint-battery4}},
            \mathrm{\eqref{constraint-load9}},
            \mathrm{\eqref{constraint-load10}},
            \mathrm{\eqref{constraint-as}},
            \mathrm{\eqref{constraint-load12}},
        \end{aligned}
    \end{equation}
where $\bm{s}_u^{\mathrm{CO}} {\triangleq} (\bm{g}^{*}_u, \bm{r}^{*}_u, \bm{l}^{\mathrm{AC} *}_u, \bm{l}^{\mathrm{FL} *}_u, \bm{c}^{*}_u, \bm{d}^{*}_u, \bm{p}^{*}_{u,v \in \mathcal{U}}, \bm{e}^{\mathrm{FIT} *}_u, \bm{e}^{\mathrm{DR} *}_u, \bm{e}^{\mathrm{AS} *}_u)$ is user $u$'s optimal energy schedule for its VPP operation. In the cooperative mode, the user $u$'s energy management solution $\bm{s}_u^{\mathrm{CO}}$ includes $\bm{p}^{*}_{u,v \in \mathcal{U}}$ that represents the optimal energy trading decisions with other users. Therefore, the user $u$ cannot locally solve the optimization problem in \eqref{opt2} because the solution $\bm{p}^{*}_{u,v \in \mathcal{U}}$ involves other users' energy usage schedules. In the next section, we will elaborate our design and implementation of the energy management algorithm to solve this problem.

\section{The Decentralized Energy Management Algorithm for Virtual Power Plant} \label{sec:solution}

In the conventional VPP system, a authorized coordinator collects all the users' energy management information and globally minimizes their operational costs in a centralized manner. Specifically, the coordinator solves the following optimization problem:
    \begin{equation}
        \begin{aligned}
            \bm{s}^{*} = & \arg \min_{ \left\{ \bm{g}_u, \bm{r}_u, \bm{l}_u^{\mathrm{AC}}, \bm{l}_u^{\mathrm{FL}}, \bm{c}_u, \bm{d}_u, \bm{p}_{u,v \in \mathcal{U}}, \bm{e}_u^{\mathrm{FIT}}, \bm{e}_u^{\mathrm{DR}}, \bm{e}_u^{\mathrm{AS}} \mid_{\forall u \in \mathcal{U}} \right\} } \sum_{\forall u \in \mathcal{U} }\mathcal{P}^{\mathrm{CO}}_u, \label{opt3}\\
            &\mathrm{subject} \: \mathrm{to} \: \mathrm{constraints} \:  \mathrm{\eqref{constraint-load5}},
            \mathrm{\eqref{constraint-load6}},
            \eqref{constraint-trading1}, \eqref{constraint-trading2},
            \mathrm{\eqref{constraint-load1}},
            \mathrm{\eqref{constraint-tmp}},
            \mathrm{\eqref{constraint-load3}},
            \mathrm{\eqref{constraint-load4}},\\
            & \mathrm{\eqref{constraint-battery1}}-\mathrm{\eqref{constraint-battery4}},
            \mathrm{\eqref{constraint-load9}},
            \mathrm{\eqref{constraint-load10}},
            \mathrm{\eqref{constraint-as}},
            \mathrm{\eqref{constraint-load12}},
        \end{aligned}
    \end{equation}
where the solution $\bm{s}^{*}$ represents the optimal energy management schedules for all the VPP users. 

However, the above conventional VPP management method has two drawbacks. First, this method incurs great privacy concerns because the users have to reveal all of their energy management information to the coordinator. The revealed information is actually the user's privacy that includes its preference on the indoor temperature ($\tau_u$) and flexible appliance ($L_u^{\mathrm{Ref}}[t]$), renewable energy generation capability ($R_u[t]$), the battery information ($\overline{B}_u, \overline{C}_u, \overline{D}_u$), and all of the user's energy usage schedule. Second, the solution $\bm{s}^{*}$ is computed by the coordinator; thus the users cannot verify and trust the result because the operation of the coordinator is a "black box" to the users. Hence, the conventional VPP management method relies on a trustable coordinator and requires established trust between the users and the coordinator. Therefore, the centralized design is vulnerable to single-point failure and incurs higher operational investment.

To address these drawbacks, we design a decentralized VPP energy management algorithm that preserves the users' privacy and also achieves the optimal energy management for all the users. Our method only requires the users to share their energy trading decisions without disclosing any other private information \rv{including grid usage, renewable energy usage,  appliance loads, battery operations, DR, FIT, and AS.}  Furthermore, by implementing the proposed energy management algorithm in \rv{the} smart contract on the blockchain, we remove the need for a centralized coordinator in the VPP system.

\begin{figure*}[!t]
    \centering
    \includegraphics[width=14cm]{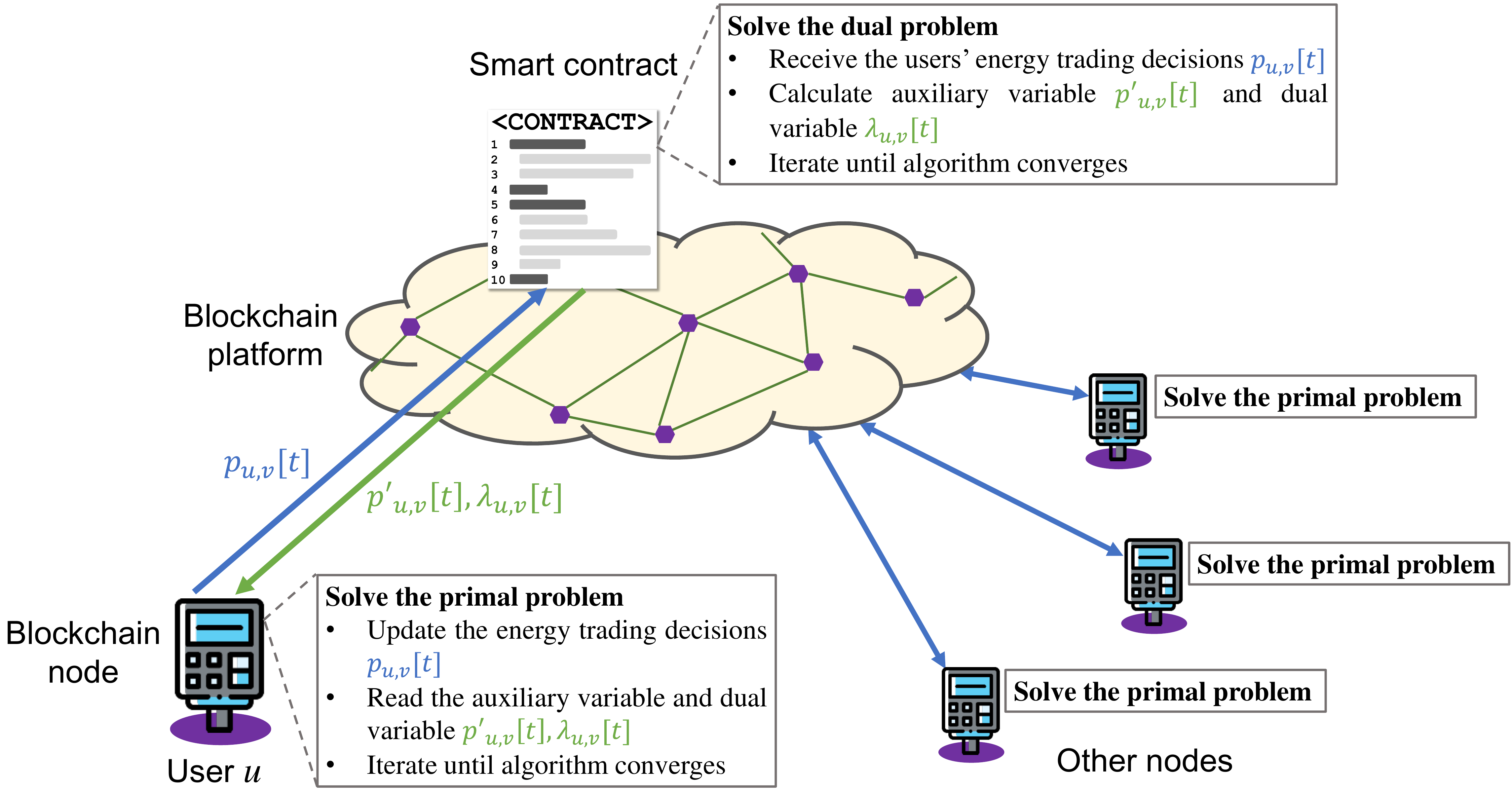}
    \caption{The process of the decentralized VPP energy management platform.}
    \label{f:process}
\end{figure*}

\subsection{The augmented Lagrangian method}

We employ the primal-dual method \citep{jakovetic2020primal} to solve the optimization problem \eqref{opt3}. According to the alternating direction method of multipliers (ADMM) method \cite{boyd2011distributed}, we first derive the augmented Lagrangian for Problem \eqref{opt3} as
    \begin{equation}
    \begin{aligned}
           \mathcal{L}  = \sum_{\forall u \in \mathcal{U} }\mathcal{P}^{\mathrm{CO}}_u
         + \sum_{u\in\mathcal{U}} \sum_{v \in \mathcal{U}} \sum_{t\in\mathcal{H}} 
        \Big[ \frac{\rho}{2} \left( {p}^{\prime}_{u,v}[t] - p_{u,v}[t] \right)^{2} & \\
         + \lambda_{u,v}[t] \left( {p}^{\prime}_{u,v}[t] - p_{u,v}[t] \right) \Big] &, \label{lagrangian}
    \end{aligned}    
    \end{equation}
where the term ${p}^{\prime}_{u,v}[t]$ is the auxiliary variable of the user $u$'s energy trading decision at time slot $t$, and $\lambda_{u,v}$ is the dual variable of ${p}^{\prime}_{u,v}[t]$. The coefficient ${\rho}/{2}$ denotes the penalty sensitivity of the auxiliary variables. 
\rv{We aim to develop a distributed optimization strategy for the energy management problem of the VPP in \eqref{opt3}, and the coupling constraints in Eq.~\eqref{constraint-trading1} and Eq.~\eqref{constraint-trading2} are the barrier to the distributed optimization. Thus, we introduce auxiliary variables to replicate energy trading decisions $\bm{p}_{u,v}$, denoted as $\bm{p}^{\prime}_{u,v} {=} \bm{p}_{u,v}$.}
Following the short notations in Tab.~\ref{tab:short}, we also define the vector form for ${p}^{\prime}_{u,v}[t]$ and $\lambda_{u,v}$ as below
\begin{align}
    &\textrm{Auxiliary variable:} \: & \bm{p}^{\prime}_{u,v} &\triangleq \Big\{ {p}^{\prime}_{u,v}[t] \mid_{\forall t \in \mathcal{H}} \Big\}, \\
    &\textrm{Dual variable:} \: & \bm{\lambda}_{u,v} &\triangleq \Big\{ \lambda_{u,v}[t] \mid_{\forall t \in \mathcal{H}} \Big\}.
\end{align}
Also, the auxiliary variable ${p}^{\prime}_{u,v}[t]$ should follow the same constraints of its original variable $p_{u,v}[t]$ in \eqref{constraint-trading1} and \eqref{constraint-trading2}, therefore we have
    \begin{align}
        p^{\prime}_{u,v}[t] & = -p^{\prime}_{v,u}[t], ~\forall t \in \mathcal{H},~\forall u,v \in \mathcal{U}, \label{constraint-auxiliary1} \\
        p^{\prime}_{u,v}[t] & = 0~\textrm{if}~u = v, ~\forall t \in \mathcal{H}. \label{constraint-auxiliary2}
    \end{align}

\subsection{The primal-dual decomposition}
Based on the augmented Lagrangian in Eq.~\eqref{lagrangian}, we decompose the original optimization problem into a primal problem and a dual problem. The primal problem is solved locally by each user; specifically, each user $u {\in} \mathcal{U}$ individually computes the following optimization problem
\begin{equation}
\begin{aligned}
    &\min \Bigg\{ \mathcal{P}^{\mathrm{CO}}_u {+} \sum_{v \in \mathcal{U}} \sum_{t\in\mathcal{H}} \Big[ \frac{\rho}{2} \left( p^{\prime}_{u,v}[t] {-} p_{u,v}[t] \right)^{2} {-} \lambda_{u,v}[t] p_{u,v}[t] \Big] \Bigg\}, \\
    &\mathrm{s.t.} \: \mathrm{constraints} \:
    \mathrm{\eqref{constraint-load5}},
            \mathrm{\eqref{constraint-load6}},
            \mathrm{\eqref{constraint-load1}},
            \mathrm{\eqref{constraint-tmp}},
            \mathrm{\eqref{constraint-load3}},
            \mathrm{\eqref{constraint-load4}},
            \mathrm{\eqref{constraint-battery1}}-\mathrm{\eqref{constraint-battery4}},\\
            & \mathrm{\eqref{constraint-load9}},
            \mathrm{\eqref{constraint-load10}},
            \mathrm{\eqref{constraint-as}},
            \mathrm{\eqref{constraint-load12}}, \\
    &\mathrm{with} \: \mathrm{variables:} \: \bm{g}_u, \bm{r}_u, \bm{l}_u^{\mathrm{AC}}, \bm{l}_u^{\mathrm{FL}}, \bm{c}_u, \bm{d}_u, \bm{p}_{u,v \in \mathcal{U}}, \bm{e}_u^{\mathrm{FIT}}, \bm{e}_u^{\mathrm{DR}}, \bm{e}_u^{\mathrm{AS}}. \label{primal}
\end{aligned}
\end{equation}

Since the objective function in Problem~\eqref{primal} is convex, the users can use standard convex optimization tool to solve the primal problem. In the primal problem, the values of the auxiliary variable ${p}^{\prime}_{u,v}[t]$ and the dual variable $\lambda_{u,v}[t]$ are calculated by the dual problem that is defined as
\begin{equation}
\begin{aligned}
    &\min \sum_{u\in\mathcal{U}} \sum_{v \in \mathcal{U}} \sum_{t\in\mathcal{H}} \Big[ \frac{\rho}{2} \left( {p}^{\prime}_{u,v}[t] - {p}_{u,v}[t] \right)^{2}
         + \lambda_{u,v}[t] {p}^{\prime}_{u,v}[t]  \Big]\\
    & \mathrm{s.t.} \: \mathrm{constraints} \:\eqref{constraint-auxiliary1}, \eqref{constraint-auxiliary2}, \\
    & \mathrm{with} \: \mathrm{variables:} \: {p}^{\prime}_{u,v}[t], \forall u,v \in \mathcal{U}, \forall t \in \mathcal{H}, \label{dual}
\end{aligned}
\end{equation}
where the value of the users' trading decisions ${p}_{u,v}[t]$ are updated by the primal problem. We derive the optimal solution of the dual problem as follows
        \begin{align}
                {p}^{\prime}_{u,v}[t] = \frac{\rho \left( {p}_{u,v}[t] {-} {p}_{v,u}[t] \right) - \left( \lambda_{u,v}[t] - \lambda_{v,u}[t] \right) }{2 \rho}, \label{updateenergy}
        \end{align}
and the dual variable $\lambda_{u,v}$ is updated by
    \begin{equation}
        \lambda_{u,v}[t] \leftarrow \lambda_{u,v}[t] + \rho \left( {p}^{\prime}_{u,v}[t] - {p}_{u,v}[t] \right). \label{updatelambda}
    \end{equation}
    
The primal-dual method works in an iterative manner. The primal problem updates the users' energy schedule decisions including the energy trading decisions ${p}_{u,v}[t]$. The dual problem uses the value of ${p}_{u,v}[t]$ to update the auxiliary variable ${p}^{\prime}_{u,v}[t]$ and the dual variable $\lambda_{u,v}[t]$, which are used by the primal problem in the next iteration. We define the sum of the Euclidean distance between the auxiliary variable $\bm{p}^{\prime}_{u,v}$ and its original variable $\bm{p}_{u,v}$ as one of the convergence criteria. We also have the difference between the dual variables over iterations $k$ and $k-1$, i.e., $\Delta \bm{\lambda} = \bm{\lambda}(k) - \bm{\lambda}(k-1)$, as the other convergence criterion, where $\bm{\lambda} \triangleq \left\{ \bm{\lambda}_{u,v} \mid_{\forall~u,v} \right\}$. The solution of the primal and dual problems will converge to the optimal solution of the original optimization problem  in \eqref{opt3} when the convergence error is small enough, so we have the following convergence conditions
\begin{equation}
    \sum_{u \in \mathcal{U}} \sum_{v \in \mathcal{U}} \parallel \bm{p}^{\prime}_{u,v} - \bm{p}_{u,v} \parallel \leq \epsilon_{1},
    ~ \parallel \Delta \bm{\lambda}  \parallel  \leq \epsilon_{2},
    \label{eq:err}
\end{equation}
where $\epsilon_{1}$ and $\epsilon_{2}$ are the convergence thresholds, which are usually set as small values.

\subsection{Algorithm implementation}\label{s:alg}
\begin{algorithm}[!tb]
     \caption{Decentralized VPP energy management algorithm.} \label{a:1} 
     \SetAlgoLined

     \KwIn{
     iteration index $k$, convergence threshold $\epsilon_{1}$ and $\epsilon_{2}$, dual variable $\bm{\lambda}^{(0)}_{u,v}$, auxiliary variable $\bm{p}^{\prime}_{u,v}$.}
     \KwOut{
     Optimal VPP energy management schedule {\small$\bm{s}_u^{\mathrm{CO}} {=} (\bm{g}^{*}_u, \bm{r}^{*}_u, \bm{l}^{\mathrm{AC} *}_u, \bm{l}^{\mathrm{FL} *}_u, \bm{c}^{*}_u, \bm{d}^{*}_u, \bm{p}^{*}_{u,v \in \mathcal{U}}, \bm{e}^{\mathrm{FIT} *}_u, \bm{e}^{\mathrm{DR} *}_u, \bm{e}^{\mathrm{AS} *}_u)$} for each user $u \in \mathcal{U}$.}
     
     $k {\leftarrow} 1$; \\
     $\epsilon_{1} {\leftarrow} 0.000001$, $\epsilon_{2} {\leftarrow} 0.000001$; \\
    $\bm{\lambda}_{u,v} {\leftarrow} \bm{0}$, $\bm{p}^{\prime}_{u,v} {\leftarrow} \bm{0}$, $\forall u,v \in \mathcal{U}$; \\
    \While{$\sum_{u \in \mathcal{U}} \sum_{v \in \mathcal{U}} \parallel \bm{p}^{\prime}_{u,v} - \bm{p}_{u,v} \parallel > \epsilon_{1}$ \rv{or $\parallel \Delta \bm{\lambda}  \parallel > \epsilon_{2}$ }
    }{
        \For{each user $u \in \mathcal{U}$}{
        $\cdot$ Read the auxiliary variable $\bm{p}^{\prime}_{u,v}$ and dual variable $\bm{\lambda}_{u,v}$ from the smart contract;
        
        $\cdot$ Solves the primal optimization problem in Eq.~\eqref{primal};
        
        $\cdot$ Updates the energy trading decisions $\bm{p}_{u,v}$ to the smart contract;
        }
    The smart contract \textbf{do}
    
    $\cdot$ Collects $\bm{p}_{u,v}$ from all the users $u \in \mathcal{U}$;
    
    $\cdot$ Computes the auxiliary variable $\bm{p}^{\prime}_{u,v}$ by Eq.~\eqref{updateenergy};
    
    $\cdot$ Computes the dual variable $\bm{\lambda}_{u,v}$ by Eq.~\eqref{updatelambda};
    
    $\cdot$ Updates $\bm{p}^{\prime}_{u,v}$ and $\bm{\lambda}_{u,v}$ to all the users;
    
    $k \leftarrow k+1$;
    }
\end{algorithm}

As discussed in the previous section, the proposed decentralized VPP energy management algorithm consists of a primal problem and a dual problem. We let the VPP users solve the primal problem in \eqref{primal} locally on their smart meters. Since \rv{the} users' primal problems are decoupled with each other, they can solve the primal problems in parallel. The solution to the dual problem is implemented on the blockchain using the smart contract.

\rv{The decentralized energy management algorithm preserves the users’ privacy by minimizing the amount of information that the user needs to reveal to other parties. The user's private information includes its grid usage ($\bm{g}_u$), renewable energy usage ($\bm{r}_u$), appliance loads ($\bm{l}_u^{\mathrm{AC}}$, $\bm{l}_u^{\mathrm{FL}}$), battery operations (charging $\bm{c}_u$, discharging $\bm{d}_u$), demand response ($\bm{e}_u^{\mathrm{DR}}$), feed-in energy ($\bm{e}_u^{\mathrm{FIT}}$), and auxiliary service ($\bm{e}_u^{\mathrm{AS}}$). As described in Algorithm~\ref{a:1}, the users’ private information is only required to solve the primal problem; therefore, the private information is well preserved because it is locally used by the smart meter and not revealed to anyone else. The only information that the users reveal is their energy trading decisions $\bm{p}_{u,v}$, which is sent to the smart contract in each iteration.}

\begin{figure*}[!t]
    \centering
    \includegraphics[width=16.5cm]{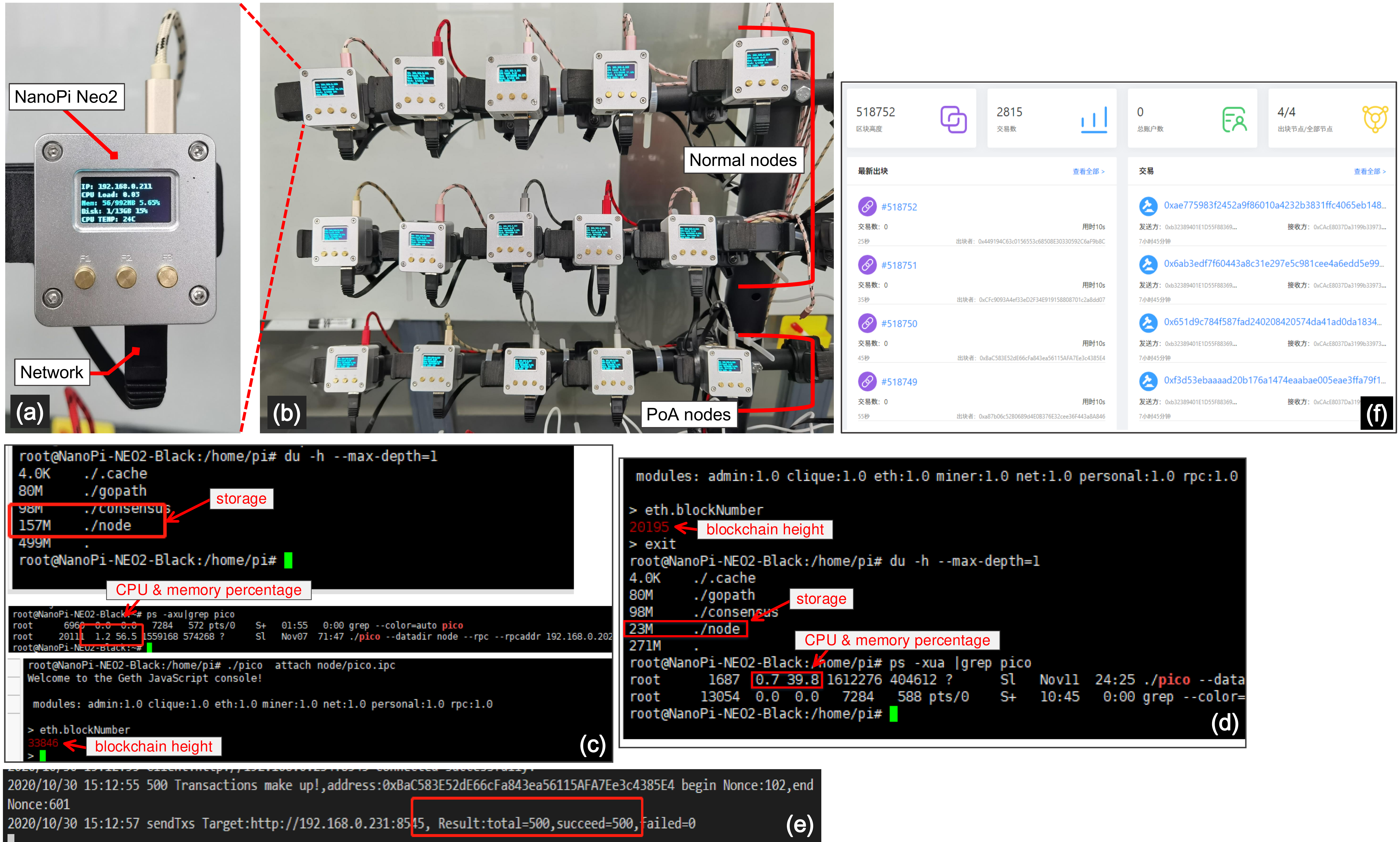}
    \caption{\rv{System implementation: (a) The NanoPi Neo2 hardware node. (b) The blockchain network that consists of 5 PoA nodes and 10 normal nodes. (c) The running status of a PoA node. (d) running status of a normal node. (e) The result of calling the smart contract. (f) The front end that shows the status of the blockchain.}}
    \label{f:hw}
\end{figure*}

The whole process of the energy management algorithm is shown in Fig.~\ref{f:process}. The algorithm iterates between the primal problem (on users' smart meters) and the dual problem (on smart contract) until it converges. In each iteration, each user solves the primal problem to obtain their own energy scheduling decisions, including the energy trading decision $\bm{p}_{u,v}$, and then update this value to the smart contract. Upon receiving the trading decision $\{ \bm{p}_{u,v}|_{\forall u,v {\in} \mathcal{U}} \}$ from all the users, the smart contract automatically solves the dual problem to obtain the auxiliary variables $\bm{p}^{\prime}_{u,v}$ and $\bm{\lambda}_{u,v}$, then update them to the users to begin the next iteration. The iteration ends when the convergence error defined in \eqref{eq:err} is below the threshold, and each user obtains its optimal energy schedule $\bm{s}_u^{\mathrm{CO}}$. Finally, all the users execute the optimal energy schedule during the VPP operation.

The primal problem is solved on the user's smart meter, which is an embedded device with limited computing resources. We numerically solve the primal problem using the quadratic programming package from the GNU Octave \citep{octave} on the Ubuntu Core operating system. The hardware specification and its performance will be discussed later in the next section.

The dual problem is solved in the smart contract that is deployed by the VPP operator on the blockchain. This smart contract is implemented in Solidity and consists of three core functions: 1) \rv{Function 1} is to solve the dual optimization problem by implementing the numerical computation of Eq.~\eqref{updateenergy} and Eq.~\eqref{updatelambda}; 2) \rv{Function 2} is to set new values to the variables that store the users' energy trading decisions $\bm{p}_{u,v}$. The users can call this function to update their local trading decisions in each iteration; 3) \rv{Function 3} is to reveal the values of the dual variables $\bm{\lambda}_{u,v}$ and the auxiliary variables $\bm{p}^{\prime}_{u,v}$ computed by the first function. The users can call this function to read the latest values of $\bm{\lambda}_{u,v}$ and  $\bm{p}^{\prime}_{u,v}$ in each iteration.

\rv{Specifically, Function 2 and 3 are simple memory access operations, which are implemented by pure Solidity programming. However, Function 1 cannot be implemented in Solidity, because the Solidity language does not support floating-point number operation, which is needed to compute Eq.~\eqref{updateenergy} and Eq.~\eqref{updatelambda}. Therefore, we implement Function 1 as a pre-compiled contract written in Go language. We integrate the code of the pre-compiled contract into the source code of the blockchain; hence, they become built-in functions of the EVM (Etherum virtual machine) that can be called in Solidity. In each iteration of the algorithm, the users interact with the smart contract by calling the three functions as described in Fig.~\ref{f:process}.
}

Algorithm~\ref{a:1} describes the working process of the proposed decentralized VPP energy management method. According to the analysis in \cite{boyd2011distributed}, this algorithm converges to the optimal solution of the centralized optimization problem in \eqref{opt3}. In this algorithm, the user's private energy schedule (including grid usage, renewable energy usage, appliance loads, DR, FIT, and AS) is processed locally by the user's smart meter; thus the user's private information is well preserved and not revealed to anyone else. The only information that the users need to reveal is their energy trading decisions $\bm{p}_{u,v}$, which is needed by the smart contract to solve the dual problem. Furthermore, the smart contract runs on the blockchain, which is a trustable computing platform, therefore removes the need for a trusted coordinator and ensures the correctness of the optimization result.

\section{System Implementation and Evaluation}\label{sec:eval}
In this section, we will present our developed blockchain-based VPP energy management platform and the simulation results for validating our algorithm and system. 

\subsection{Blockchain-based VPP energy management platform}\label{s:blockchain}
We present the blockchain-Based VPP energy management platform, including the setup of the experiment, blockchain system, and the algorithm implementation.

\subsubsection{Experiment setup}

To evaluate the feasibility and performance of the proposed blockchain-based VPP energy management platform, we build a proof-of-concept network of smart meters as shown in Fig.~\ref{f:hw}b. Because we cannot access the operating system of a real smart meter, we use the NanoPi~Neo2 \citep{nanopi} to emulate the hardware of the smart meter. The NanoPi~Neo2 is a low-power embedded ARM board with an ARM-A53 (quad-core 1.5GHz) CPU and 1GB memory. We package the board with a metal housing, a 16GB SD card, and an OLED screen to display the device information as shown in Fig.~\ref{f:hw}a. We use an Ethernet router to connect the 15 NanoPis to form a small-scale private network for our experiment. All the NanoPis run the Ubuntu Core 16.04 as the operating system.

\subsubsection{The blockchain system}
We adopt the Ethereum \citep{eth} as the underlying blockchain system in this work based on the following considerations. First, Ethereum is a mature and well-verified blockchain system that has been applied in many areas. Second, Ethereum supports the smart contract that is needed to implement the Algorithm~\ref{a:1}. Third, Ethereum is an open-source project written in Go Language so that we can modify its source code to tailor the blockchain to our application scenario.

To run the Ethereum blockchain on the smart meter, we modified the consensus protocol from PoW (proof of work) to PoA (proof of authority). The original Ethereum uses the PoW consensus protocol that is computation-intensive and memory-intensive. In our test, running a PoW node on the NanoPi consumes 100\% of the quad-core CPU time and all of its memory, which heavily slows down the operating system. By contrast, the PoA consensus protocol consumes much fewer hardware resources, thus is more suitable for smart meters. \rv{In our test, we observed that a PoA node consumes about 1.2\% CPU time, 565MB memory, and 157MB storage (when block height is 33846) as shown in Fig.~\ref{f:hw}c; and a normal node consumes about 0.7\% CPU time, 390MB memory, and 23MB storage in Fig.~\ref{f:hw}d. We also found that the memory consumption of the blockchain node can be further reduced by decreasing the number of concurrent TCP links. These hardware requirements can be satisfied by inexpensive embedded devices and modern smart meters \citep{nxp}.}

To build a test network of the blockchain system, we let five NanoPis be the PoA nodes and let the other ten NanoPis be the normal nodes (Fig.~\ref{f:hw}b). We use the wireless router to limit the network bandwidth of the NanoPis to 250Kbps, which is close to the network condition of the practical smart meter network. In the experiments, the measured average throughput of the blockchain is about 200 TPS (transactions per second), and the observed peak throughput is 780 TPS. The smart contract (in Solidity) is also supported by the Ethereum virtual machine (EVM) module. \rv{A simple web-based front end is built to show the status of the blockchain in Fig.~\ref{f:hw}f.} The experimental result shows that the underlying blockchain is sufficient to execute the proposed decentralized VPP energy management algorithm.

\subsubsection{The implementation of Algorithm~1}
To verify the feasibility of the proposed method, we implement the solution of the primal problem on the NanoPi using GNU Octave. After solving the local primal problem, the NanoPi sends transactions to interact with the smart contract as described in Section~\ref{s:alg}. Fig.~\ref{f:hw}e illustrates the execution result of the smart contract. In the test, we set the value of the convergence threshold $\epsilon_{1}, \epsilon_{2}$ to $0.000001$. Our Algorithm 1 converges within $40$ iterations with the testing data, which shows that the proposed method is feasible in practice.

\subsection{Performance evaluation}
\begin{figure}[!t]
    \centering
    \includegraphics[width=8.5cm]{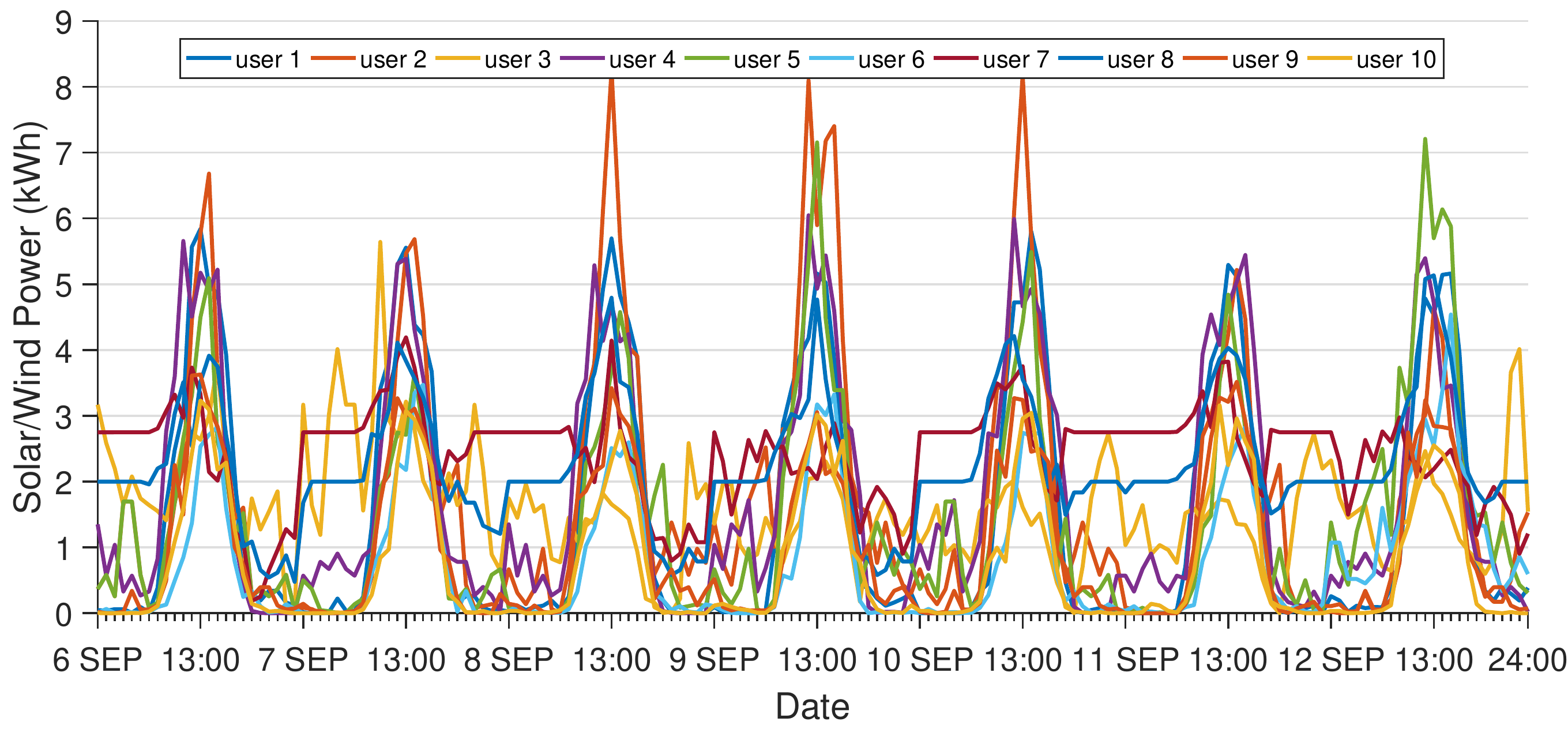}
    \caption{The ten users' renewable energy generation capacity during the simulation period.}
    \label{f:re}
\end{figure}

\begin{figure}[!t]
    \centering
    \includegraphics[width=8.5cm]{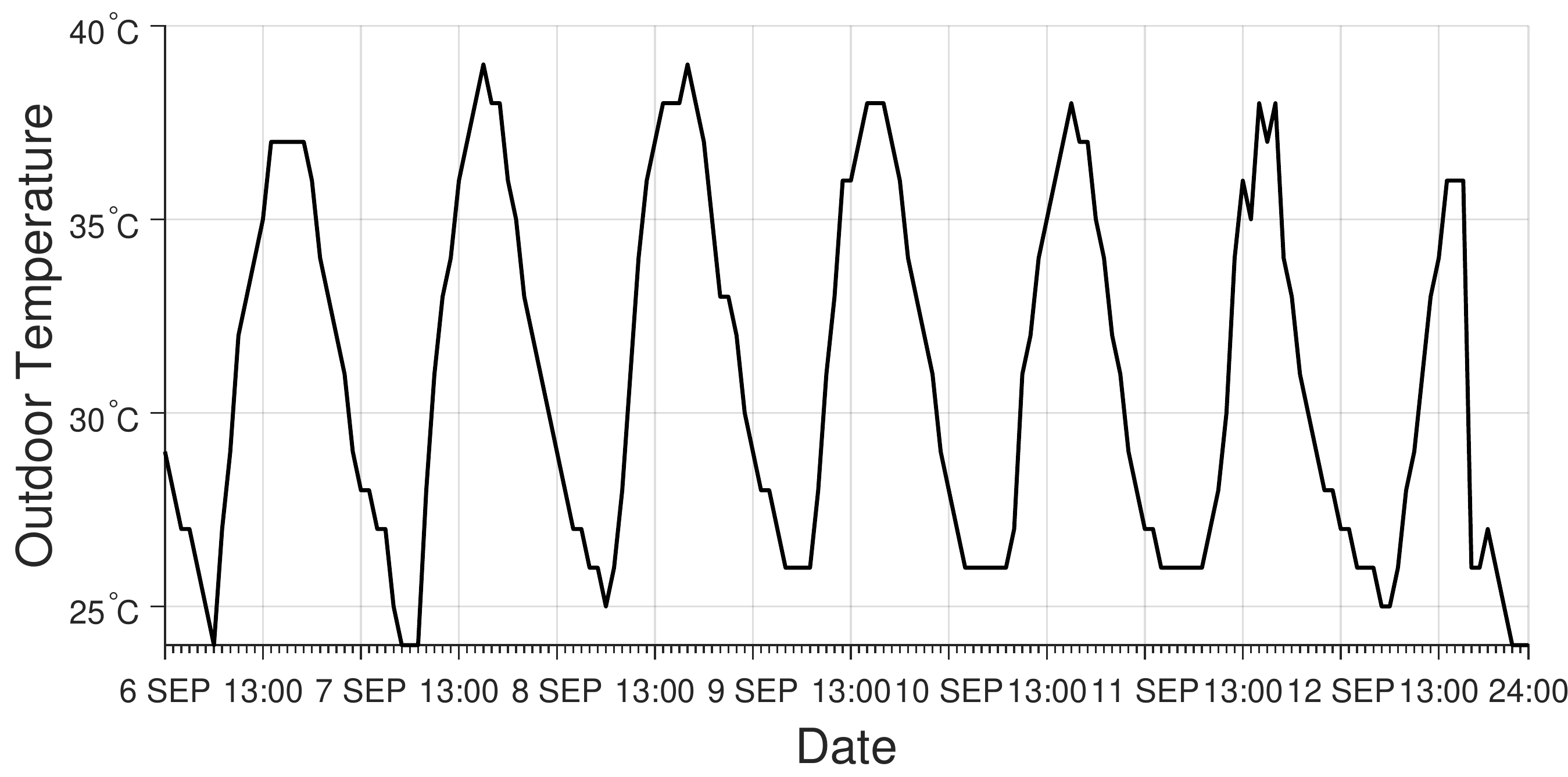}
    \caption{The outdoor temperature in the simulation period of one week.}
    \label{f:tmp}
\end{figure}

To evaluate the performance of the proposed Algorithm~\ref{a:1}, we collect some energy usage data from real-world smart grid system \citep{wang2015joint, pecan}, including solar/wind power generations shown in Fig.~\ref{f:re}, in-house energy consumption records \citep{pecan}, and outdoor temperature shown in Fig.~\ref{f:tmp}. The users' battery capacities are randomly generated in the range from 10kWh to 15kWh in the simulation. Note that, in Fig.~\ref{f:re}, the solar power generation is mainly active during the daytime, and the wind power generation lasts longer and is more stable. We use these data as the inputs of Algorithm 1 (described in Section~\ref{sec:solution}) to determine the optimal day-ahead VPP energy schedule. 

\begin{figure*}[!t]
    \centering
    \includegraphics[width=14.5cm]{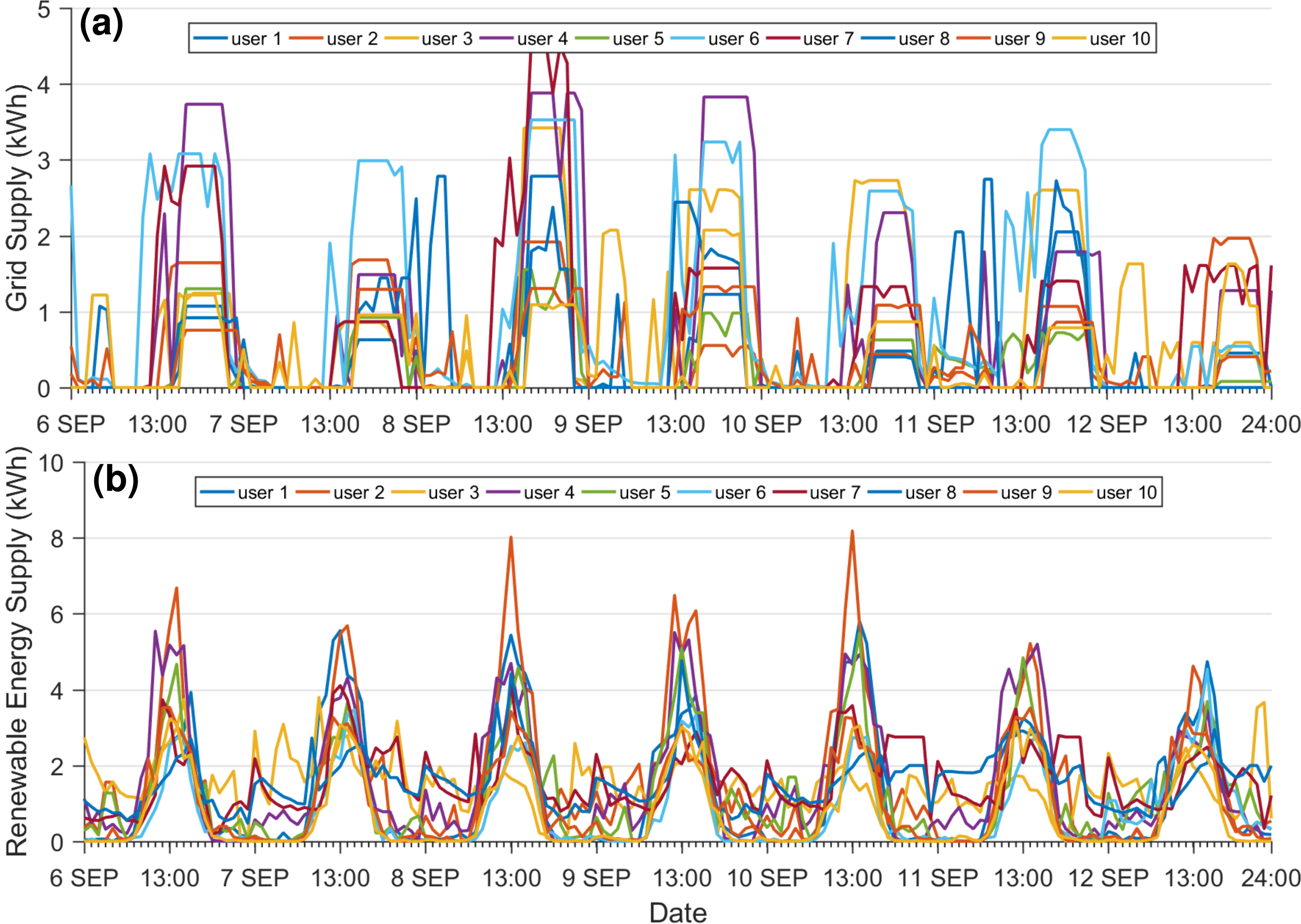}
    \caption{The ten users' optimal power supply schedules with the decentralized VPP energy management algorithm: (a) grid supply and (b) renewable energy supply.}
    \label{f:supply}
\end{figure*}

\subsubsection{Power supply scheduling in decentralized VPP} 
We then implement the decentralized VPP energy management method (Algorithm~\ref{a:1}) in Matlab and using the previously described simulation data to evaluate its performance. The algorithm's outputs are the users' optimal VPP energy schedules for the next day, as the day-ahead energy scheduling method is used in this work. Our simulation window is one week (168 hours based on simulation data during September 6-12), and the user's energy schedules include the usage of (adjustable and flexible) appliances, power supplies (from grid and renewable), P2P energy trading, and network services (such as feed-in tariff, demand response, and ancillary service).

Fig.~\ref{f:supply} shows the optimal energy schedule of all the ten users' power supply, including the grid supply in Fig.~\ref{f:supply}(a) and the renewable energy supply in Fig.~\ref{f:supply}(b). From the grid supply usage in Fig.~\ref{f:supply}(a), we observe that the grid supply and the renewable energy are well complemented, which shows the proposed VPP energy management algorithm reduces the users' dependence on the grid. Furthermore, the flat plateaus in Fig.~\ref{f:supply}(a) show that the two-part tariff pricing scheme of the grid effectively achieves peak-shaving on the users' grid power usage. Comparing Fig.~\ref{f:supply}(b) with Fig.~\ref{f:re}, we see that the users' renewable energy \rv{is} well utilized by the distributed management algorithm. 

\begin{figure*}[!t]
    \centering
    \includegraphics[width=14.5cm]{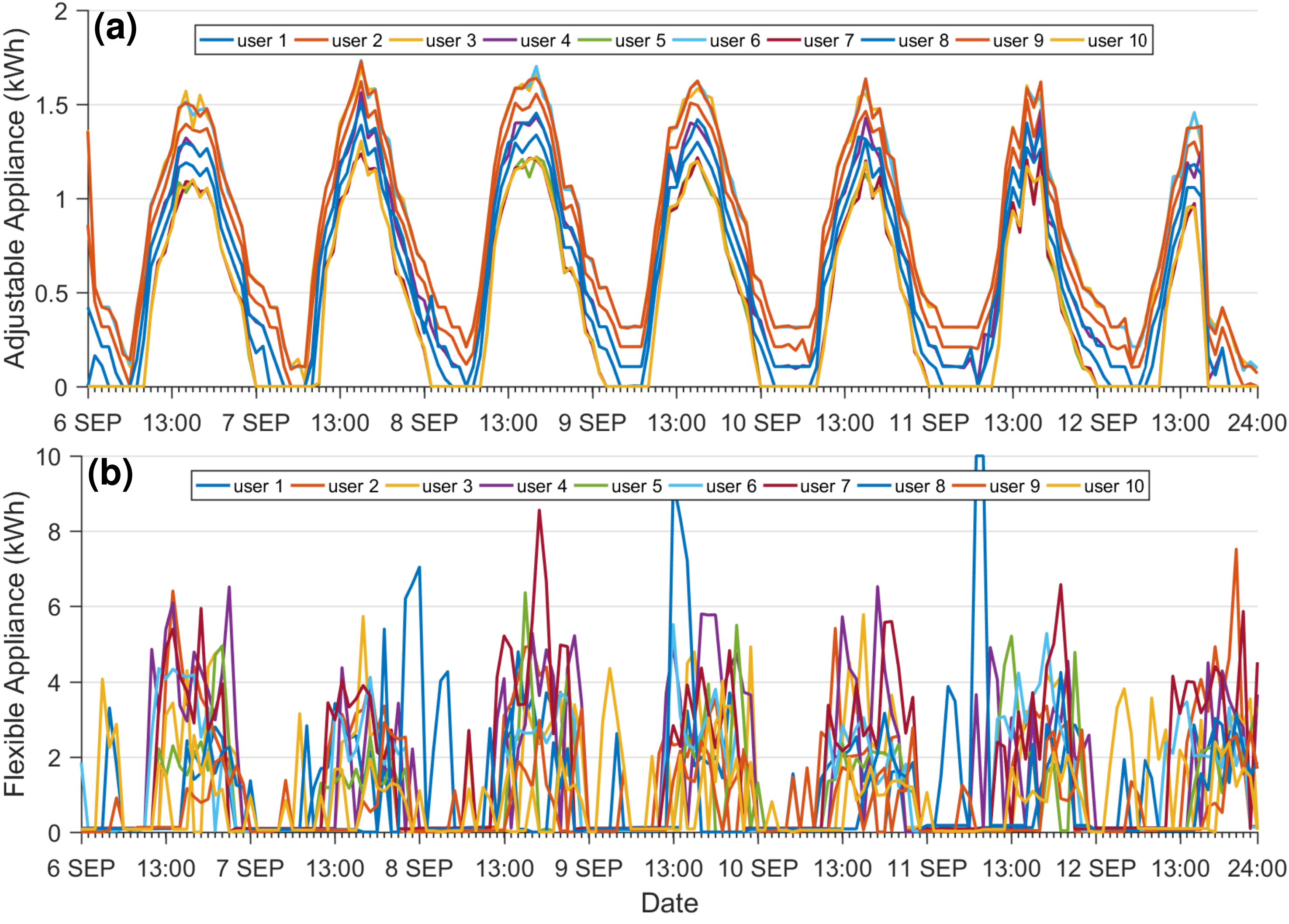}
    \caption{The ten users' optimal appliances usage schedules with the decentralized VPP energy management algorithm: (a) schedule of adjustable appliances, and (b) schedule of the flexible appliances.}
    \label{f:load}
\end{figure*}

\begin{figure*}[!t]
    \centering
    \includegraphics[width=14.5cm]{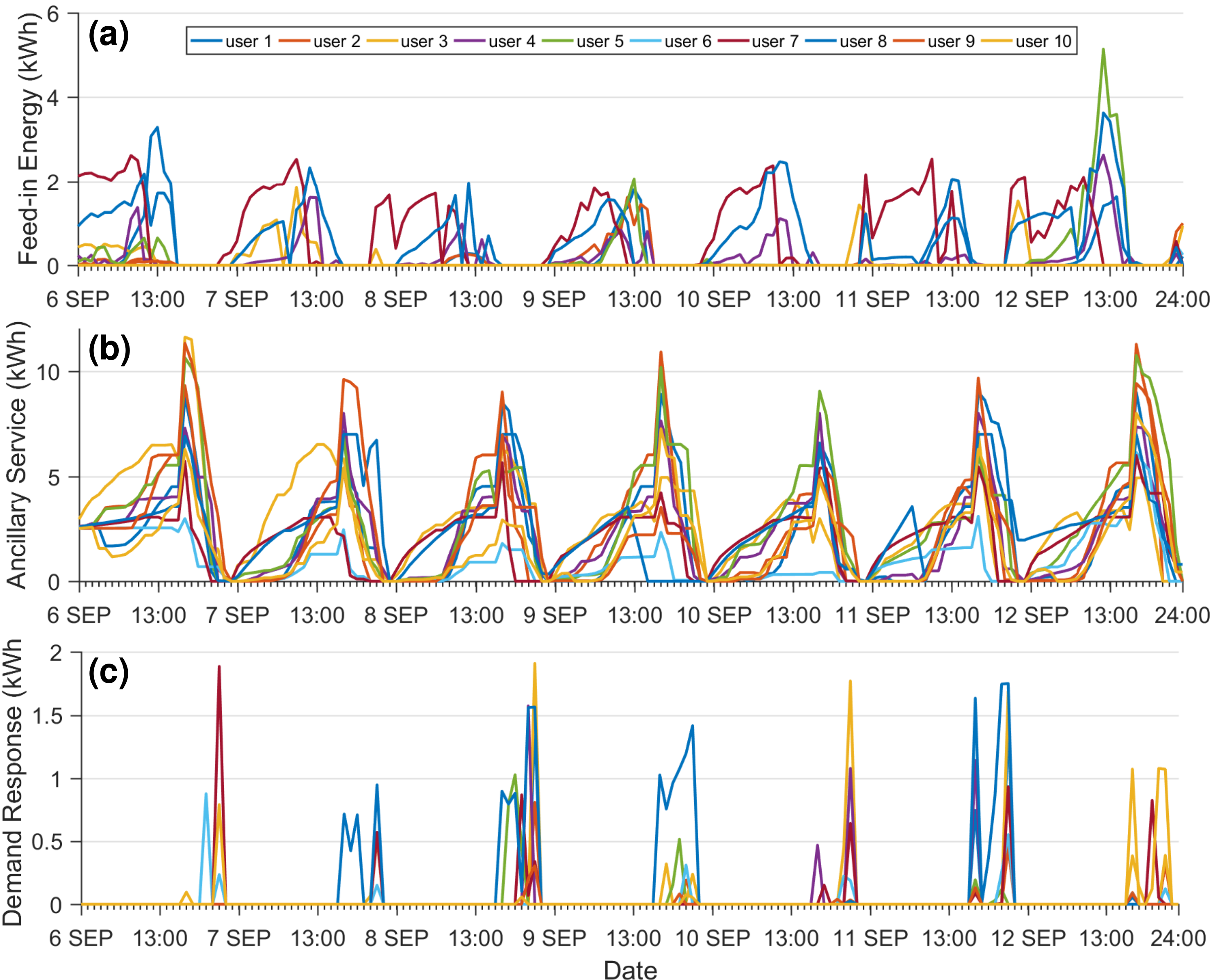}
    \caption{\rv{The ten users' network services scheduled by the decentralized VPP energy management algorithm: (a) feed-in tariff; (b) ancillary service; (c) demand response.}}
    \label{f:vpp}
\end{figure*}

\begin{figure*}[!t]
    \centering
    \includegraphics[width=14.5cm]{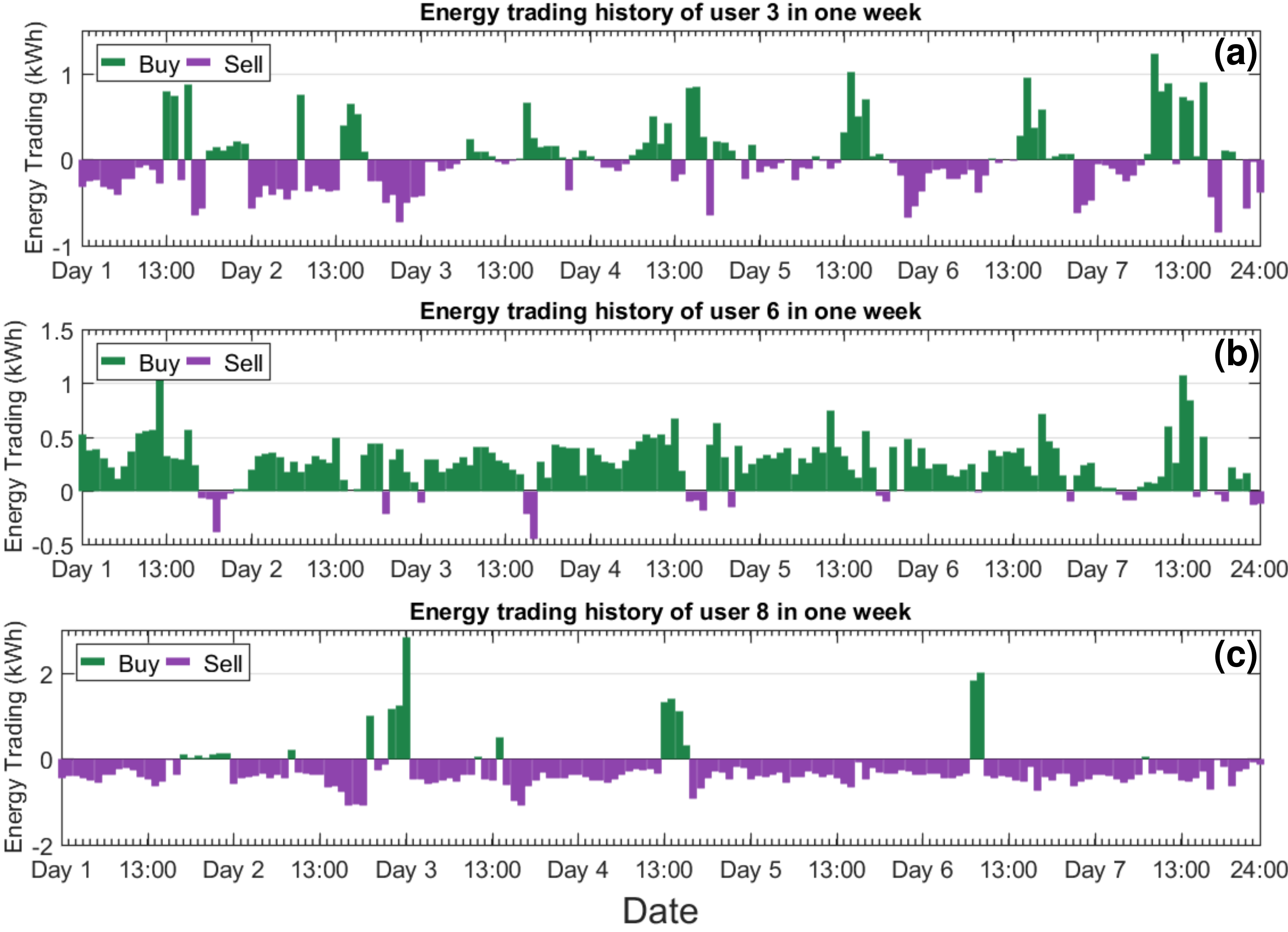}
    \caption{The optimal energy trading schedule of three users (a) user 4, (b) user 6, and (c) user 7. Note in this figure a positive value means buying energy and a negative value means selling energy.}
    \label{f:et}
\end{figure*}

\subsubsection{Electrical appliance scheduling}
Fig.~\ref{f:load} shows the optimal scheduling of users' appliance load, including adjustable appliance load in Fig.~\ref{f:load}(a) and flexible appliance load in Fig.~\ref{f:load}(b). We see that users' adjustable appliance loads all follow the trend of the outdoor temperature shown in Fig.~\ref{f:tmp} but exhibit differences as a result of different preferred indoor temperature settings. The flexible appliance loads show very devise patterns. Because each user has its own local renewable generations and preference for energy trading and network services, leading to uniquely best scheduling of its flexible appliance load.

\subsubsection{Network services}
Network services aggregate users' energy resources and provide multiple revenues to users. Fig.~\ref{f:vpp} illustrates the optimized network services that users determine to provide, including feed-in energy in Fig.~\ref{f:vpp}(a), ancillary service in Fig.~\ref{f:vpp}(b), and demand response in Fig.~\ref{f:vpp}(c). We see from Fig.~\ref{f:vpp}(a) that users sell extra renewable generations to the grid when their demand is low. In Fig.~\ref{f:vpp}(b), users provide reserve services using idle capacities of batteries when they do not need to charge and discharge, especially during the afternoon they have charged enough from the renewable but do not have a high demand until the evening. Users also perform demand response in Fig.~\ref{f:vpp}(c) when they are called to lower the consumption from their adjustable appliances during peak hours in the evening. 

\subsubsection{P2P energy trading}
Fig.~\ref{f:et} depicts the optimal energy trading profiles of three typical users over one week. We see that the energy trading profiles of three users (user 3, user 6, and user 8) are complementary. User 3 often needs to buy energy in the daytime but \rv{sells} extra energy at night. User 6 and user 8 exhibit opposite energy-trading patterns, in which user 6 always needs to buy energy while user 8 has a lot of extra energy to sell. Our developed blockchain-based energy management platform provides opportunities for users to interact with each other to exploit their diverse generation and load profiles and thus gain benefits, such as reduce their costs. 

\subsubsection{Cost reduction and payment}

\begin{figure*}[!t]
    \centering
    \includegraphics[width=14.5cm]{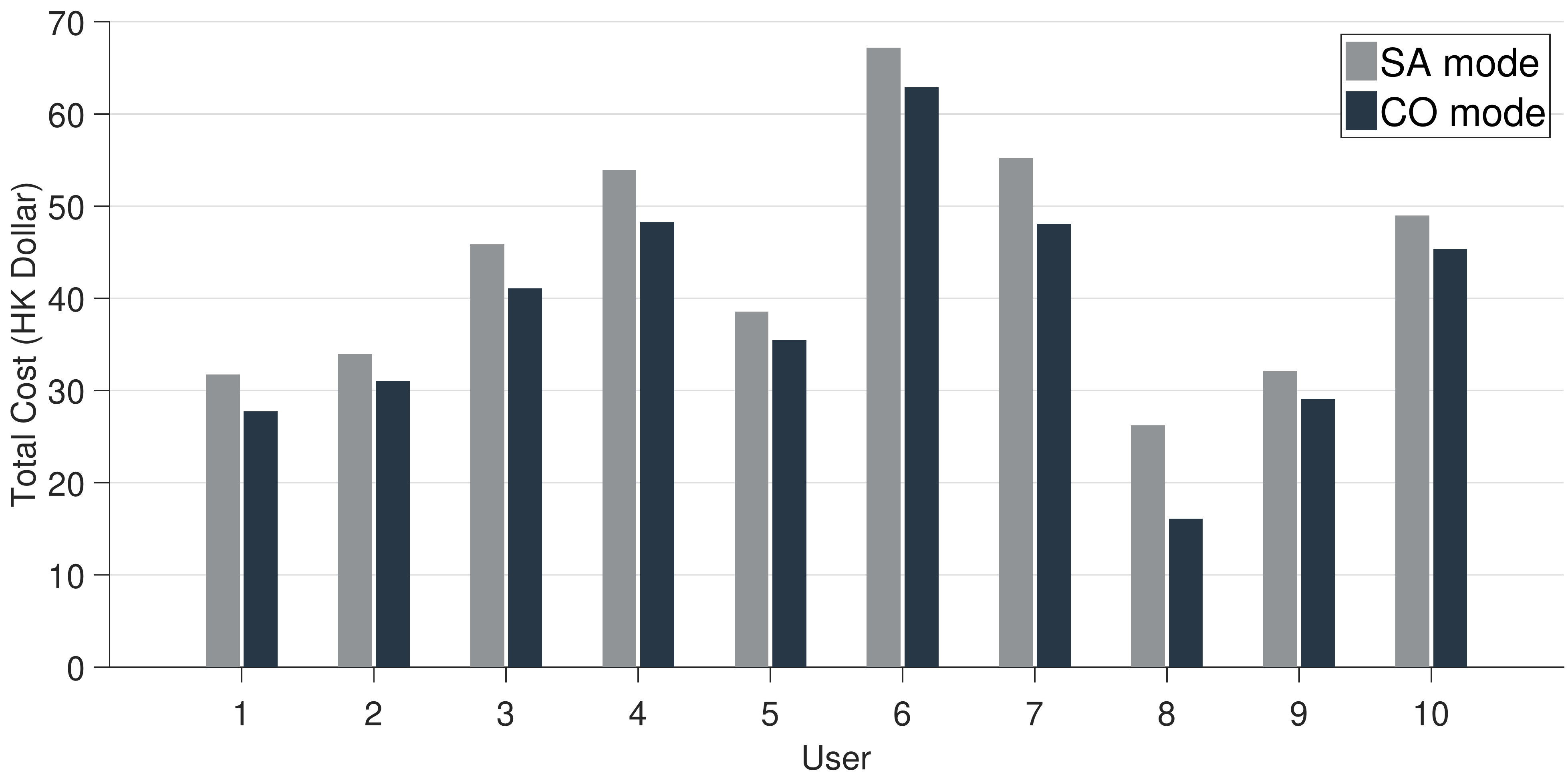}
    \caption{\rv{The comparison of the users' total costs in the two VPP modes. SA is the standalone VPP energy management without energy trading; CO is the cooperative VPP energy management with energy trading.}}
    \label{f:overall_cost}
\end{figure*}

The comparison of users' total costs is depicted in Fig.~\ref{f:overall_cost}, showing the costs under both SA mode and CO mode. In the SA mode, all the users independently schedule their energy usage without energy trading, and thus the costs are higher. In the CO mode, users can interact with each other to trade energy following Algorithm 1, and thus their costs are reduced. By employing the distributed energy management algorithm, the cost reduction of user 1 to 10 are listed in Table~\ref{t1:costreduction}, and the sum cost of all the users are reduced by 11.2\%.

\begin{table}[htb]
    \centering
    \renewcommand{\arraystretch}{1.1}
    \caption{The comparison of all users' costs in the two modes.}
    \label{t1:costreduction}
    \begin{tabular}{c c c c}
        \hline
        \multirow{2}{*}{User} & \multicolumn{2}{c}{Total Cost \rv{(HK Dollar)}} & \multirow{2}{*}{Reduction (\%)} \\
        \cline{2-3}
        & SA mode & CO mode &   \\
        \hline
        1 & 31.7 & 27.7 &  12.5\% \\
        2 & 33.9 & 31.0 &  8.6\% \\
        3 & 45.8 & 41.1 &  10.3\% \\
        4 & 53.9 & 48.3 &  10.4\% \\
        5 & 38.5 & 35.4 &  8.0\% \\
        6 & 67.1 & 62.8 &  6.3\% \\
        7 & 55.2 & 48.0 &  12.9\% \\
        8 & 26.2 & 16.1 &  38.6\% \\
        9 & 32.0 & 29.1 &  9.2\% \\
        10 & 48.9 & 45.3 &  7.4\% \\
        \hline
    \end{tabular}
\end{table}

\newpage
\section{Conclusion and future work}\label{sec:conclusion}
This paper developed a blockchain-based virtual power plant energy management platform, including distributed energy trading algorithm design and blockchain system implementation. Specifically, we modeled energy trading and network services for residential users with various loads, energy storage, and local renewables. The users can interact with each other to trade energy and choose to provide network services aggregated through the VPP. Given users' independence, we designed a distributed optimization algorithm to manage the energy schedules, energy trading, and network services of users. We also developed a prototype blockchain system for VPP energy management and implemented our algorithm on the blockchain system. We validated our blockchain-based VPP energy management platform through extensive experiments and simulations using real-world data. The simulation results showed that our designed algorithm effectively manages the users' energy schedules, energy trading, and network services in the VPP and also demonstrated the effectiveness of our blockchain system.

For our future work, we will further improve the energy management platform by removing any centralized communication and computation to make the VPP more decentralized and flat. We will also consider and test a larger system with hundreds or thousands of users in the VPP by designing more efficient distributed algorithms.

\section{ACKNOWLEDGMENTS}
This work is in part supported by the National Natural Science Foundation of China (project 61901280) and the FIT Academic Staff Funding of Monash University.

\bibliography{mybibfile}

\end{document}